\documentclass{article}
\usepackage[utf8]{inputenc}
\usepackage{longtable}
\usepackage{units}
\usepackage{amsmath}
\usepackage{amssymb}
\usepackage{graphicx}
\usepackage{esint}
\usepackage{nicefrac}
\usepackage{float}
\usepackage{siunitx}
\usepackage[version=3]{mhchem}
\usepackage{authblk}

\newcommand{\latin}[1]{{\it #1}}

\author{Ivan Kryven \thanks{i.kryven@uva.nl\\
		  Tel.:~+31 20 525 6423\\
          Fax.:+31 20 525 5604\\}}
\author{Jorien Duivenvoorden}
\author{Joen Hermans}
\author{Piet D.Iedema}

\affil{University of Amsterdam,\\ Science Park 904,  Amsterdam, The Netherlands}

\title{Random graph approach to multifunctional molecular networks}

\begin{document}

\maketitle
\begin{abstract}
Formation of a molecular network from multifunctional precursors is modelled with a random graph process. 
The random graph model favours reactivity for monomers that are positioned close in the network topology, 
and disfavours reactivity for those that are obscured by the surrounding. The phenomena of conversion-dependant reaction rates, gelation, and micro-gelation are thus naturally predicted by the model and do not have to be imposed.
Resulting non-homogeneous network topologies are analysed to extract such descriptors as: size distribution, crosslink distances, and gel-point conversion. 
Furthermore, new to the molecular simulation community descriptors are invented. These descriptors are especially useful for understanding evolution of pure gel, amongst them: cluster coefficient, network modularity, cluster size distribution.
\end{abstract}

\begin{figure}[H]
\begin{center}
Table of Contents graphic\\
  \includegraphics[width=0.8\textwidth]{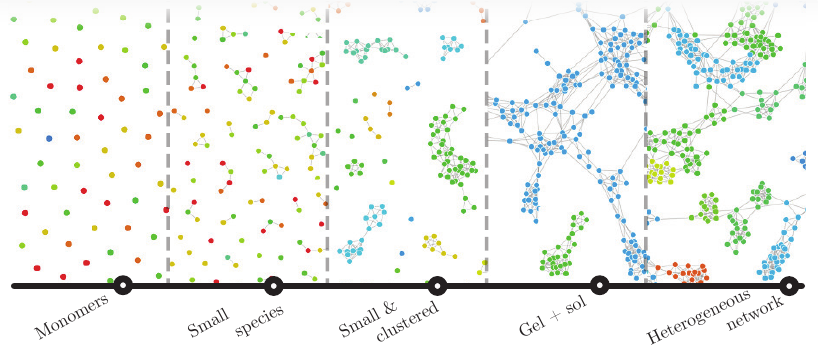}
\label{default}
\end{center}
\end{figure}

\section*{Nomenclature}
\begin{longtable}[l]{ll}
$\alpha$ & 	constant related to chain stiffness and volume exclusion\tabularnewline
$a'$  & scaling factor for average shortest path\tabularnewline
$\boldsymbol{A}$  & adjacency matrix\tabularnewline
$b_{k}$  & disconnected components of the network\tabularnewline
$c_{e},\;c_{i},\;c_{m}$  & reaction propensity coefficients \tabularnewline
$c_{k},\;c(A)$  & local and global clustering coefficients\tabularnewline
$d_{i}$  & degree of node $i$\tabularnewline
$d_{i,\text{free}}$  & free degree of node $i$\tabularnewline
$d_{i,\text{max}}$  & maximum degree of node $i$\tabularnewline
$E_{\text{max}}$  & maximum number of edges in the network\tabularnewline
$f_{i}$  & concentration of fatty acids with $i$ unsaturations\tabularnewline
$g_{i}$  & hindrance effect \tabularnewline
$g_{01}(\chi)$  & gel indicator function\tabularnewline
$H(\chi)$  & Heaviside function\tabularnewline
$k_{p},\;k_{c},\;k_{m}$  & reaction rate constants \tabularnewline
$l$  & average length of a chemical bond \tabularnewline
$n$  & number of nodes in the network \tabularnewline
$n_{A}$  & Avogadro number\tabularnewline
$N$  & number of samples in MC\tabularnewline
$O(n)$  & big O notation for asymptotic behaviour of a function\tabularnewline
$p_{i,j}$  & shortest path between nodes $i,j$ \tabularnewline
$p$  & average shortest path \tabularnewline
$P(A)_{i,j}$  & probability of the next reaction between monomers $i,j$ at given topology $A$\tabularnewline
$P^{*}(A)_{i,j}$  & proposal probability for the MCMC scheme\tabularnewline
$\mathcal{P}(k)$  & Poisson distribution\tabularnewline
$Q$  & stochastic matrix for computing steric hindrance\tabularnewline
$R$  & effective reaction distance \tabularnewline
$\rho$  & molar density\tabularnewline
$s_{1}(q),\;s_{2}(q)$  & Markov chain for MCMC sampling\tabularnewline
$S_{1}(n),\;S_{2}(n)$  & size of largest, and second largest components for an $n-$monomer system\tabularnewline
$t$  & reaction time\tabularnewline
$\tau$  & waiting time between reactions\tabularnewline
$u_{i}$  & uniformly distributed random variable\tabularnewline
$U[0,1]$  & uniform distribution\tabularnewline
$V$  & volume\tabularnewline
$W_{n}$  & concentration of TAGs of functionality n\tabularnewline
$\Phi(\boldsymbol{A})_{i,j}$  & cyclisation rate for nodes $i,j$\tabularnewline
$\chi$  & conversion of edges\tabularnewline
$\chi_{g}$  & Conversion at gel transition\tabularnewline
$z$  & current number of edges in the network\tabularnewline
\end{longtable}

\section{Introduction}

Polymer science profits from the rapid development in the computing
field in a very straightforward way: the complexity of macromolecular
architecture or geometry is, to a growing extent, captured by simulation
of the most basic physical processes. Instead of following heuristics
leading to the classical mathematical equations. We anticipate that
properties of continuum can be obtained as emerging from repeated
simulations of the many-body physics at the nano-level, a strategy
that seems to be more properly placed in stochastic simulations than
in deterministic mathematical modelling. That said, one must realise
that such simulations imply an enormous spread of length and time
scales: even the simple case of a linear polymer exhibits geometrical
structure from the scale of a chemical bond (1A) to the scale of the
gyration radius (100A), and collective length scales in dense materials
often are even much larger. Moreover, our drive for modelling is not
limited to linear polymers but also is directed at tree-like branched
structures, cross-linked polymers, and even infinite molecular networks,
\latin{i.e.} \emph{gels}. In this paper we propose a strategy that draws a
consensus between what can be computed on the one hand and pursuing
fundamental roots of the matter at hand on the other.

The process of assembly or polymerisation of molecular networks usually
goes through a few temporal stages: disconnected monomers, linear
polymer, branched polymer, and gel (an infinite network) \cite{dusek2012,dusek2013, kryven2014a}. Furthermore, the gel itself should
not be perceived as a single, final state of polymer topology. After
the gel point the connectivity patterns of the network continue to
evolve\cite{akagi2009,zhou2012,armitage2000}. In fact, in many cases
a major part of the whole process of conversion may occur in the gel
regime\cite{kryven2014b}. 

Side by side with new developments in Polymer Reaction Engineering,
a novel field of \emph{network science} has been introduced recently.
In a much broader sense, the main object of studies in network science
is networks regardless their background, \latin{i.e.} not necessarily molecular
networks. In chemistry, not surprisingly as chemical bonding leads
to connectivity problems, one of the main toolboxes in network science,
graph theory is already in use, for instance, in the form of the Wiener
index \cite{rouvray2002}. However, apart from this line of research,
the applicability of graph theory to large or infinite molecular networks, has
received little attention. The current paper presents a cross dissemination
between polymer science and network science as disciplines, with special
care devoted to polymer-related issues. The problem we deal with in the present paper has three main features.\\ 
\emph{a)}
The functionality of the monomer is limited and satisfies a predefined
frequency distribution. The monomer functionality is crucial for the
gel formation and properties of the material\cite{schwenke2011,lange2011,lang2003}. A bi-functional monomer just forms linear chains. Monomers with higher functionality may lead to gels or infinite networks. Star
macro-monomers may carry many functional groups, thus being a perfect
base for complex molecular networks.\\
 \emph{b)} The average reaction rates may decrease by orders of magnitude
due to diffusion limitation as a consequence of the decrease of the
free volume caused by the formation of the polymer network, \latin{e.g.} as in Ref. \cite{goodner2002}. The Trommsdorff or gel effects in radical polymerisation
is an example of this. Moreover, the mobility of the larger molecules
is affected more strongly than that of small molecules. This may affect
the rates of reaction, depending on the type of molecules involved
in those reactions. The issue of 'chain length dependent termination'
is a frequently studied example of this phenomenon \cite{kattner2015}.
Furthermore, due to steric hindrance effect some monomers might be
less available for the reaction being shielded by the surrounding
topology of the network. The available polymerisation models accounting
for shielding\cite{eichinger2005} have a mean-field nature and consequently
do not allow for correct interpretation of possible inhomogeneities
or irregularities present in network topology. \\
\emph{c)} The topological
distance between reacting units has to be accounted for. The distance
between reacting units is not an issue in systems where these units
make part of separate larger assemblies moving so fast that their
average distance - or concentration - is the dominating factor. In large
 networks, however, the units are relatively fixed and morel likely to react if the distance is small.  
In atomistic simulations position and reactivity are addressed
by the 'capture sphere growth approach'\cite{eichinger2005}.
In percolation theory monomer position and reactivity is described using a grid in 3D space.
However, even without performing simulations in actual three-dimensional space, it is possible to incorporate
geometrical information inferred from the network topology \cite{kryven2013c},
from analytical results on random walks\cite{dusek2014,wang1952,dusek1978},
or from population balance equations \cite{kryven2014g,kryven2014d}. 

Here, we present a new approach based on Random Graph Modelling, which
must be considered as the first attempt ever in non-atomistic modelling,
to account for all the three features described above. It is based
on the kinetic Monte Carlo concept\cite{wang2009a,wang2009b,lazzari2014}, where we explicitly deal with position-reactivity by using the graph-theoretical
concept of the shortest path between monomer pairs. This is one significant
step further than recent kinetic MC modelling studies, where the impact
of size of individual (polymer) molecules on reactivity is accounted
for \cite{mastan2015}. In this paper we will show that this is an ambitious
but still feasible approach, as it is not computationally very demanding.
Given the fact that kinetic MC typically requires $10^9$
to $10^{11}$ units to achieve statistically relevant
results, one might expect that introducing the three features described
above would lead to excessively expensive calculations, but we will
show that this is not the case. It is indeed the higher functionality
of the monomers that creates a computationally very favourable
condition concerning the required size of the domain simulated, typically
$10^4$ units. 

This paper is structured as follows. To introduce the idea of application
Random Graph Models to polymerisation problems, we explain the key
concepts, such as: adjacency matrix, relation between topological
and spatial distances, relation between steric hindrance and node
centrality. Then, this concepts are used to build a Gillespie's stochastic
simulation algorithm that describes evolution of network topology
from monomers up to the full conversion of bonds. We also explain
how the stochastic simulations can be accelerated applying the Markovian
Chain Monte Carlo (MCMC) principle. Finally, the theoretical results
are fortified with an extensive study on a realistic system: polymerisation
of triacylglycerides, which is more commonly known as the 'drying'
of oil paint. Simulation results include widely used in the reaction
engineering community descriptors as, for example, size distribution,
gel point conversion, distance between crosslinks. Furthermore, novel
descriptors for molecular networks that are especially suitable for
post gel region, are introduced. Among these descriptors: modularity
of network, clustering coefficient, average path, and cluster-size distribution.

\section{The model system}

\paragraph{Star macromonomers}

\begin{figure}
\center \includegraphics[width=0.7\textwidth]{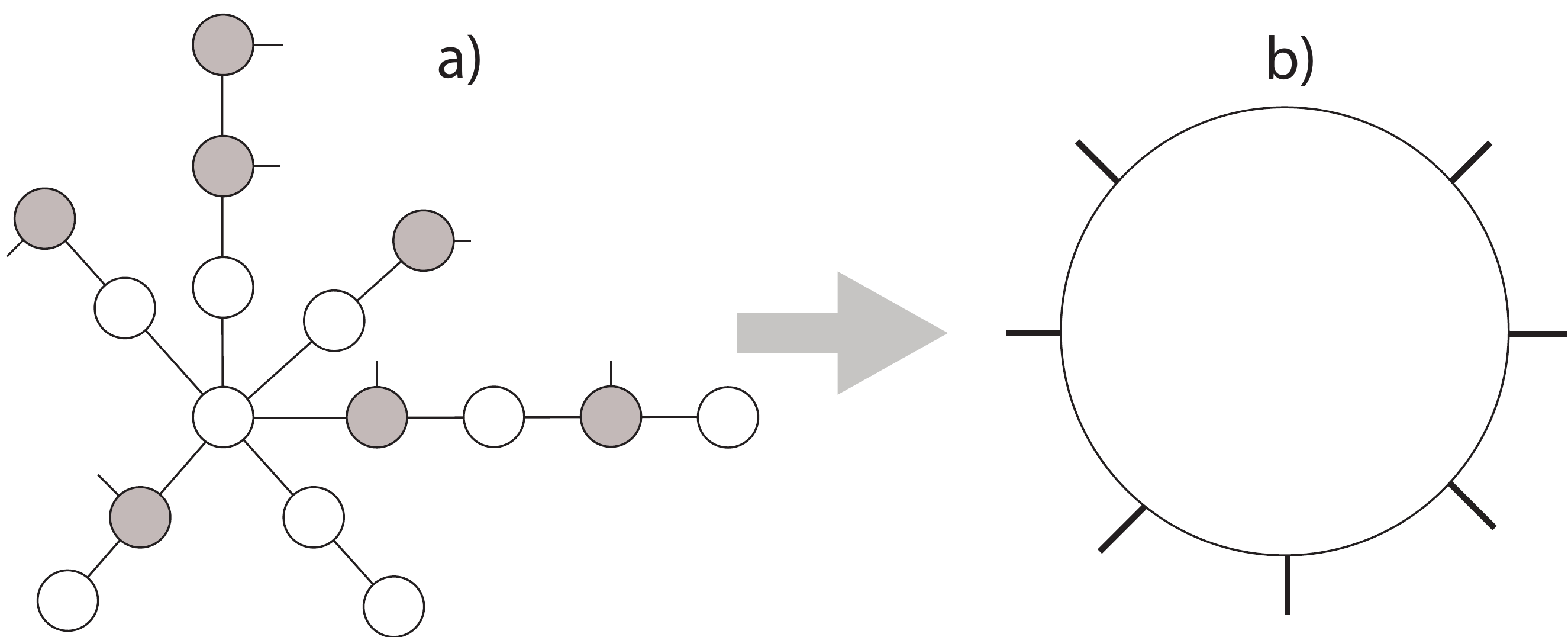} \caption{\emph{a)} A schematic representation for the multifunctional macromonomer
as a star polymer molecule. Shaded nodes carry free functional groups
and thus provide means for further polymerisation into a complex network.
\emph{b)} Coarse representation for a monomer as a 'blob' retaining
the number of functional groups. }
\label{fig:star} 
\end{figure}

Star polymers consist of a few arms joined together on one end \cite{schwenke2011}.
The arms may carry multiple functional groups, (Figure~\ref{fig:star}),
making star polymers a very attractive basis for construction of larger
macromolecular networks. In the current paper we develop a model for
polymerisation of macromonomers that are essentially star polymers
of various functionality. Since in this study the focus is on the
connectivity patterns of such monomers incorporated into a network,
we will use terms monomer and node interchangeably, depending on the
chemical or topological nature of the issue at hand. The monomers
will be distinguished according to their functionalities $d_{i,max}.$
A frequency distribution of functionalities is assumed to be known.
\begin{figure}
\center \includegraphics[width=0.6\textwidth]{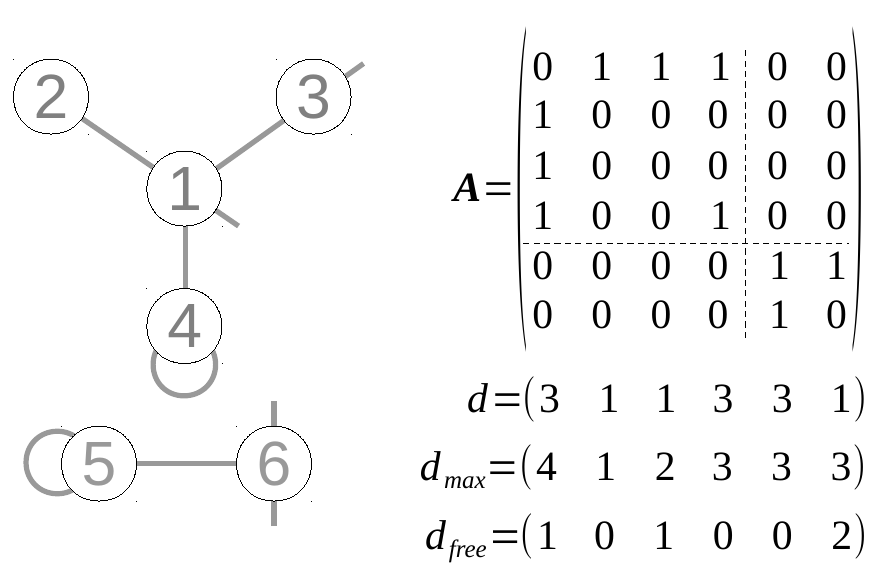} \caption{ An example of a network model with 6 nodes, the corresponding adjacency
matrix $\boldsymbol{A},$ and maximum degree vector $d_{max}$. There
are six nodes in the system: nodes 4 and 5 have a loop; nodes 3,4,5
have all functionalities used, $d_{i,free}=0;$ nodes 1,3,6 have free
functionalities available. The adjacency matrix $\boldsymbol{A}$
is block diagonalised to emphasise the fact that there are two connected
components in the system. The maximum number of edges is $E_{max}=16,$
and the current number of edges used is 12, or as a conversion fraction,
$\chi=\frac{3}{4}$. }
\label{fig:A} 
\end{figure}

\paragraph{Graph representation of a molecular network}

In chemical context, graphs have been used to represent connectivity
of (carbon) atoms in an organic molecule with a priori defined structure\cite{janezic2015,rouvray2002}.
Here, we represent the whole molecular system as a single undirected
graph, regardless of the system composition: as disconnected monomers,
ensemble of polymers, or a single connected network. Nodes numbered
with index $i=1,\dots,n$ represent monomers, and edges represent
chemical bonds between such monomers. Each node can have at most $d_{i,max}$
edges. The adjacency matrix of the molecular network is defined as
a square matrix, 
\begin{equation}
\boldsymbol{A}_{i,j}=\begin{cases}
1, & \text{node }i\text{ has bond with node }j,\\
0, & \text{no bond between }i\text{ and }j
\end{cases}
\end{equation}
The degree of node $i$ may be lower then its maximum degree, so we
distinguish between the maximum degree $d_{max}$ (monomer functionality
defined beforehand), the actual degree, 
\[
d_{i}(\boldsymbol{A})=\!\sum _{1\leq i\leq j\leq n}\!\!\boldsymbol{A}_{i,j}+\boldsymbol{A}_{i,i},
\]
and the free degree, 
\[
d_{i,free}=d_{i,max}-d_{i}(\boldsymbol{A}).
\]
The adjacency matrix $\boldsymbol{A}$ together with the maximum degree
$d_{i,max}$ constitute the model representation for a state of a
particular molecular system. Figure~\ref{fig:A} gives a simple example
of a 6-monomer system and the corresponding matrix representation.
In view of limitations posed on node's degree, $d_{i,max},$ the
maximum number of edges in the whole network is also limited, $E_{max}\leq0.5\sum_{i}d_{i,max}.$
The last equation becomes equality if $d_{i,max},\;i=1,\dots,n$ is
a \emph{graphic sequence}\cite{eggleton1979}, namely: $\sum_{i}d_{i,max}$
is an even number, and for any $r\leq n-1,$ the following inequality
holds, 
\begin{equation}
\sum _{i=1}^{n}d_{i,max}<r(r-1)+\sum _{i=r+1}^{n}\min(r,d_{i,max}).
\end{equation}
The graphic sequence is an important concept for finite simulation
systems. Since the degrees of nodes are completely random (up to a
specific frequency distribution), it is useful to constrain the degree
sequence of a finite simulation system to be graphic and thus eliminating
small-size system effects that are not in an infinite network. This
idea may be compared to imposing periodic boundary conditions in a
simulation box for partial differential problems in order to avoid
the effect of boundary layers.

Knowing a maximum number of edges allows defining a natural way of
quantifying the polymerisation progress that evolves from $0$ to
$E_{max}$ edges, 
\begin{equation}
z=\sum _{i=1}^{n}d_{i}(\boldsymbol{A}),\;z=0,\dots,E_{max},
\end{equation}
or as a relative conversion $\chi=\frac{z}{E_{max}}\in[0,1].$ Before
proceeding to defining the principle of choosing a pair of nodes receiving
a new edge on each polymerisation step, we need to introduce two
concepts first: the idea of distance between nodes, and a measure
of the node's centrality, which will allow us to address steric hindrance
effects.

\paragraph{Topological distance}

The probability of a reaction between two monomers incorporated in
the network is smaller if they are positioned far apart in space.
Our model does not explicitly include spatial configurations, but
certain information on the distance between any two nodes is encoded in
the connectivity pattern defined by the adjacency matrix $\boldsymbol{A}$.
The shortest path from node $i$ to node $j$ is the shortest sequence
of intermediate nodes that link $i,j$ together. The length of the
shortest path is the number of edges in this sequence. If nodes $i,j$
are adjacent then the shortest path is 1; if they do not belong to
the same connected component then the path is defined to have length
0. While the shortest path $p_{i,j}$ does not explicitly tell us
how far the nodes are in space, it sets an upper bound on the distance:
the two nodes cannot be further apart then the shortest path times
length of the edge, $l$. Later on, we will exploit this fact to develop
a notion for network density. Let $p_{i,j}$ denote the shortest path
between nodes $i$ and $j$. One can always find such a path by applying
a path-finding algorithm directly on the adjacency matrix $A$, for
instance the Dijkstra algorithm\cite{jungnickel2008}. In a finite
network, the average shortest path, 
\[
p=\frac{1}{n\cdot(n-1)}\cdot\sum_{i\ne j}p_{i,j}
\]
is proportional to the radius of gyration and hence there is a proportionality
between the volume occupied by the monomer network and the volume
of the gyration sphere: $\rho n=V\propto p^{3}$. Assuming a homogenous
density of the network $\rho$, augmenting the number of nodes from
$n$ up to $a\cdot n$ will cause a scale-up of the average shortest
path to $a'\cdot p$. Th exact value of $a'$ depends on spatial configuration
of the monomers but the lower bound is given by 
\begin{equation}
a'\geq\frac{\left(\frac{4}{3\pi}a\rho n\right)^{1/3}}{\left(\frac{4}{3\pi}\rho n\right)^{1/3}}=a^{1/3}\label{eq:scaling}
\end{equation}
Thus, any polymerisation model has to respect the scaling ratio $\frac{a'}{a^{1/3}}\geq1$,
otherwise the produced topologies will not have any feasible spatial
configurations. Even though this condition is necessary but not sufficient,
it allows us to restrict ranges for input parameters for the model,
post factum. 
\begin{figure}
\center \includegraphics[width=1\textwidth]{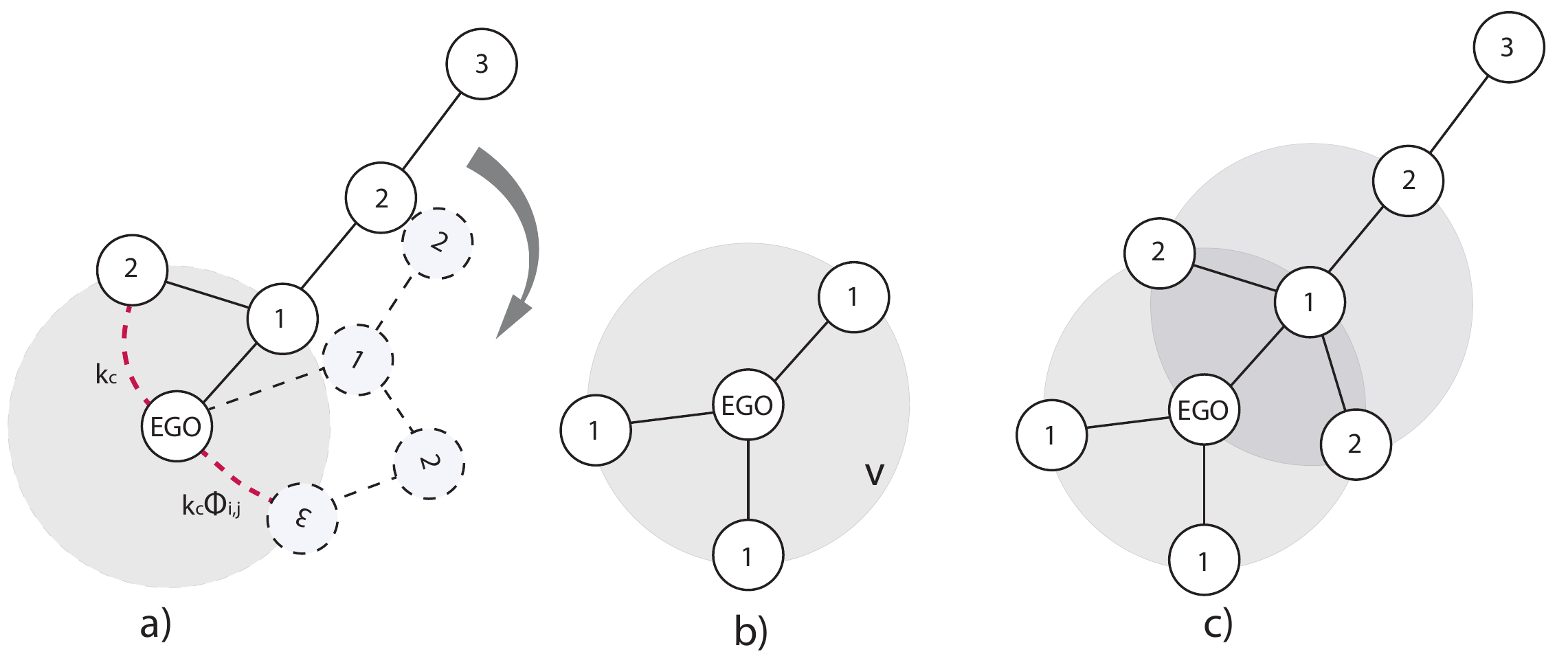}
\caption{ Network topology as seen from the perspective of a single node (marked
as 'EGO'). Numbers indicate the shortest path distance to EGO. a)
The reaction probability of two monomers incorporated in the network
depends on the probability of the path configuration that brings these
nodes together (b-c). Both the immediately surrounding and the more
remote nodes contribute to the steric hindrance effect. }
\label{fig:EGO} 
\end{figure}

In order to formulate the reaction rate between two nodes incorporated
in the network ($p_{i,j}>0$), we will follow a similar reasoning.
This time we additionally account for spatial configurations, assuming
that the shortest paths can be mimicked by self-avoiding random walks.
Two monomers need to be at minimal reaction distance $R$ to allow
reaction. For a sub-chain approximated by a random walk\cite{lang2005},
the probability of two ends being at distance $R$ is proportional
to 
\begin{equation}
\Phi^{*}(A)_{i,j}\left(\frac{3}{2\pi l^{2}\,p_{i,j}}\right)^{3/2}e^{-\frac{3R^{2}}{2l^{2}p_{i,j}}}.\label{eq:32}
\end{equation}
Now, let us estimate the probability that a random walk is self-avoiding
given that its end-to-end distance is $R$. We assume that all monomers
are distributed uniformly in the the gyration volume $V=ap^{3/2}.$
Here $a$ is a constant related to the stiffness of a polymer chain.
Since the positions of all monomers $x_{k}=0,\dots,x_{p}$ are distinct,
the probability that position $x_{k}$ does not coincide with all
previous positions is $1-k\,ap^{-3/2}.$ Hence, the total probability
that the whole chain is self avoiding is given by 
\[
\prod _{k=1}^{p}(1-k\,a\,p^{-3/2})\approx e^{-a\,p^{3/2}\sum _{k=1}^{p}k}=e^{-0.5\,a\,p^{-3/2}\,(1+p)p}\approx e^{-\alpha p^{-1/2}}.
\]
Note that a similar argument has been proposed by Flory\cite{flory1949}.
Flory, furthermore, derived the value of $\alpha$ from measurable physical
constants related to solvent properties, the stiffness of a chain,
and monomer volume. Combining the last relation with \eqref{eq:32}
and taking $R=l$ gives the probability of a self-avoiding cyclic
chain configuration, 
\begin{equation}
\Phi(\boldsymbol{A})_{i,j}=Cp_{i,j}^{-3/2}e^{-\frac{3}{2}p_{i,j}^{-1}-\alpha p_{i,j}^{1/2}}.\label{eq:Phi}
\end{equation}
The constant $C$ is chosen to satisfy $\Phi(2)=1$ allowing to relate
the intramolecular reaction rate constant to the propagation rate
constant, as will be shown later on. Another constant appearing in
\eqref{eq:32}, $\alpha,$ defines the rate of decline of the reaction
probability, asymptotically at large distances. Note that $\Phi(\boldsymbol{A})_{i,j}=0$
means that these nodes do not belong to the same connected component
and that their spatial distance cannot be inferred from the topology.
In this case the reactivity will be defined by a different mechanism
based on the chemical rate constant and the degree of steric hindrance.

\paragraph{Steric hindrance as a centrality measure}

Here, we will construct a centrality measure, defined by numbers $g_{i}\in[0,1]$
that will reflect how much a node is obscured by other nodes in the
network, and thus experience a steric hindrance effect. Matrix function
$g(\boldsymbol{A})$ is set up according to three principles: \\
 1) the nodes of higher degree are more obscured: if $d_{i}>d_{j}$
then $g_{i}<g_{j}$;\\
 2)if the node is not obscured then $g_{i}=1$; $g_{i}=0$ means full
obscuration;\\
 3) from two nodes of the same degree, the node adjacent to a more
obscured one is more obscured.\\
 In order to satisfy the first principle it is sufficient to choose
$g_{i}$ proportional to $1/d_{i},$ which also has a logical interpretation
from the excluded volume point of view (Figure~\ref{fig:EGO}b). The
second principle might be achieved by an appropriate rescaling. It
is only the third principle that requires to incorporate information
on the specific topology $A$ (Figure~\ref{fig:EGO}c). Altogether,
the principles can be reformulated in the form of an implicit equation
for the reciprocal $g_{i}^{*}=\frac{1}{g_{i}}$, 
\begin{equation}
g_{i}^{*}=\lambda_{i}d_{i}\sum _{j\sim i}g_{j}^{*},\label{eq:gi}
\end{equation}
where the summation is performed over all nodes $j$ adjacent to node
$i$, and $\lambda_{i}$ is a normalisation factor required to ensure
existence of the solution. The implicit equation \eqref{eq:gi} rewritten
in a matrix form yields an eigenvalue problem, 
\begin{equation}
\begin{aligned}g^{*}= & \;\,Qg^{*},\\
Q_{i,j}= & \;d_{i}\sum _{j}\frac{A'_{i,j}}{\sum _{k}A'_{k,j}d_{j}}
\end{aligned}
\label{eq:eig}
\end{equation}

\begin{figure}
\center \includegraphics[width=0.7\textwidth]{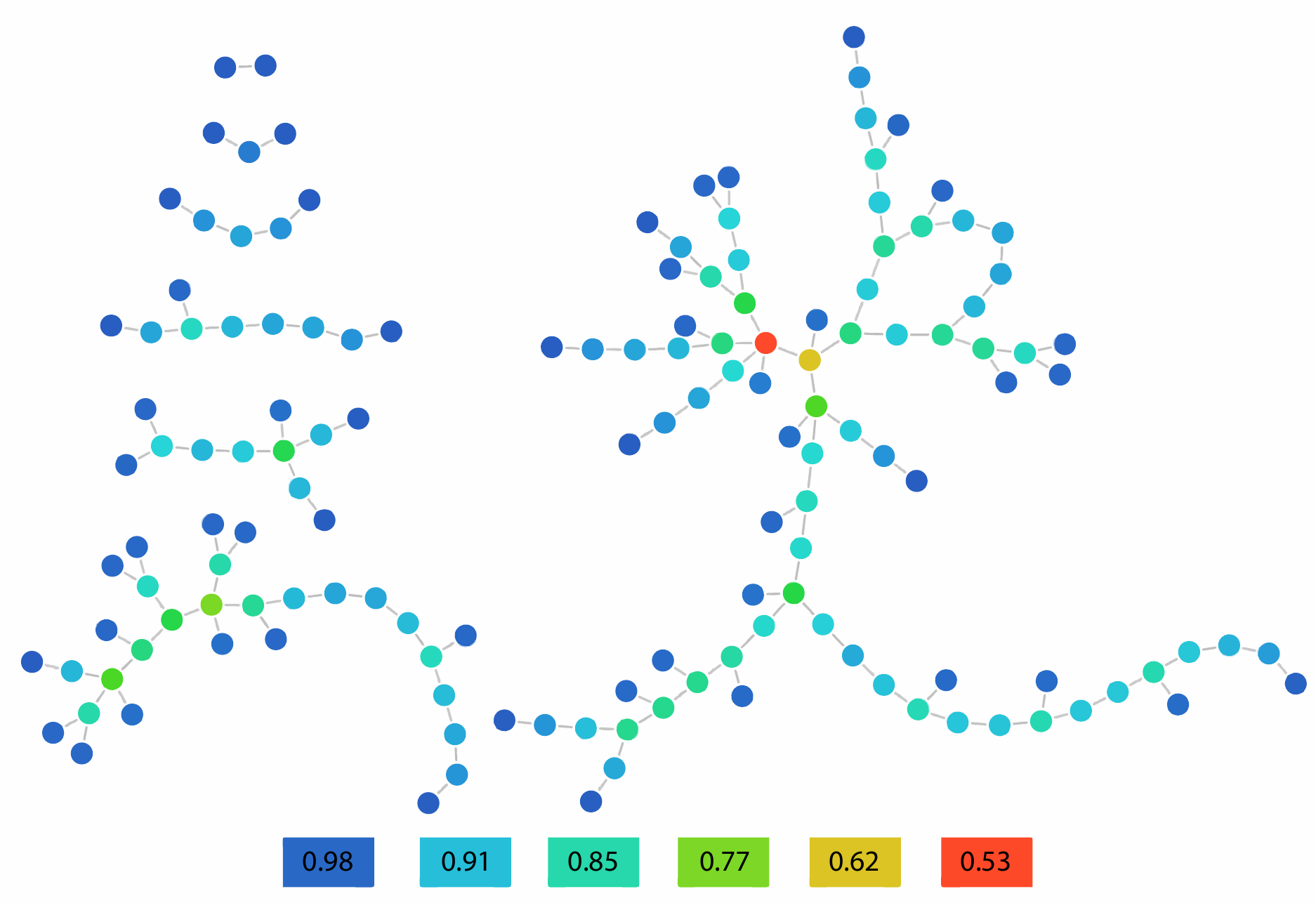} \caption{ A sample of a network topology with values of steric hindrance factors,
$g_{i}(A),$ indicated for each node. The example shows that $g_{i}(A)$
is defined by both the degree of a node and its position in the topology. }
\label{fig:Centrality} 
\end{figure}

Here matrix $\boldsymbol{A}'$ coincides with $\boldsymbol{A}$ except
for the diagonal where $(\boldsymbol{A}')_{i,i}=1.$ Matrix $Q$ has
the same size and number of zero elements as the adjacency matrix
A. With no loss in generality we may consider $d_{i}>0$ for all nodes
(otherwise $d_{i}=0$ would yield a trivial solution $g^{*}=g_{i}=1$
due to the second principle). The eigenvalue problem as formulated
in \eqref{eq:eig} is ready to use in the case of a fully connected
network (\latin{i.e.} gel). In the case of a few network components that are
not connected to each other, $\boldsymbol{A}$ has block-diagonal form.
Solving the eigenvalue problem in each block of matrix $\boldsymbol{A}$
would lead to a correct ranking within each separate component, but
it would not allow to relate the extent of steric hindrance of two
different components. To overcome this problem, we normalise $g$
in each component of the network (each block of block-diagonalised
$\boldsymbol{A}$): 
\[
g_{i}=\frac{\min\limits _{i\in b_{k}}(g_{i}^{*})}{g_{i}^{*}},\;\;i\in b_{k},
\]
where $k=1,\dots,m$ counts connected components $b_{k}$ of the network
$\boldsymbol{A}$. An example of a network topology with $g_{i}$
computed for each node is given in Figure~\ref{fig:Centrality}.

The fact that the matrix $Q$ in the eigenvalue problem \eqref{eq:eig}
is normalised, allows to alternatively view $Q$ as a Markov (left-stochastic)
matrix defining a random walk on the network\cite{newman2010}. Recall
that $Q_{i,j}>0$ if and only if $A_{i,j}=1,$ hence Q defines a weighted
graph with the same edges as in $A$. From this perspective, we may
view the solution $g_{i}^{*}$ as a frequency node-visiting by the random walk on the weighted graph $Q$. This means that instead
of solving the expensive eigenvalue problem \eqref{eq:eig}, it is
sufficient to consider a fixed point iteration $g^{*}=Qg^{*}$ starting
with an arbitrary vector with all non-zero components . 

\section{Stochastic simulation}

We consider the process of network formation (or polymerisation) as
a stochastic process $\boldsymbol{A}(z),\;z=0,\dots,E_{max}$ starting
from an empty adjacency matrix $\boldsymbol{A}(0)$ and taking it to a matrix with
one edge $\boldsymbol{A}(1)$, and so on up to, finally, $\boldsymbol{A}(E_{max}).$
This evolution proceeds by means of sampling steps: matrix $\boldsymbol{A}(z+1)$
is obtained from matrix $\boldsymbol{A}(z)$ by placing an additional
edge. Generally speaking, there are $E_{max}-z$ alternatives to decide
where to put a new edge. This choice is performed according to a conditional
probability accounting for all possibilities for the new edge to appear,
obeying the reaction propensity coefficients $c_{e},c_{i},c_{m}$
and given the topology $\boldsymbol{A}$, 
\begin{equation}
P(\boldsymbol{A})_{i,j}=\lambda_{z}\cdot\left\{ \begin{aligned} & \,c_{e}\,d_{i,free}d_{j,free}\Big(g_{i}(\boldsymbol{A})g_{j}(\boldsymbol{A})\Big)^{\beta} & i\neq j, & \;\Phi(\boldsymbol{A})_{i,j}=0;\\
 & \,c_{i}\,d_{i,free}d_{j,free}\Phi(\boldsymbol{A})_{i,j}, & i\neq j, & \;\Phi(\boldsymbol{A})_{i,j}>0;\\
 & \,c_{m}\,\binom{d_{i,free}}{2}, & i=j.
\end{aligned}
\right.\label{eq:P}
\end{equation}
Here the normalisation coefficient $\lambda_{z}$ is selected to satisfy
\begin{equation}
\sum _{i,j}P(\boldsymbol{A}(z))_{i,j}=1.\label{eq:lambda}
\end{equation}
The first line of \eqref{eq:P} refers to a new edge between nodes
forming distinct components of the network (in the case of an intermolecular
reaction, no path exists between $i,j:$ $\Phi(\boldsymbol{A})_{i,j}=0$).
The probability $P(\boldsymbol{A})$ here is defined via the node's
degree $d_{i}$ and its steric hindrance factor $g_{i}$. The intermolecular
propensity constant $c_{e}$ is related to dimerisation rate, $c_{e}=\frac{2k_{p}}{Vn_{A}},$
where $V=\frac{n}{n_{A}\rho}$ denotes the reaction volume, and $k_{p}$
is the usual rate constant for the reaction, 
\[
P_{1}+P_{1}\xrightarrow{k_{p}}P_{2},
\]
where $P_{1}$ denotes monomer, and $P_{2}$ dimer. The second line
of \eqref{eq:P} refers to a new edge between nodes from the same
connected component (in case of an intramolecular reaction, a path
between $i,j$ exists: $\Phi(A)_{i,j}>0$). The probability $P(\boldsymbol{A})$
incorporates the free degree of the nodes $d_{i}$ and the probability
that $i,j$ are positioned close in space, $\Phi(\boldsymbol{A}).$
The intramolecular propensity constant $c_{i}$ is not dependent on
volume this time, but it is directly equal to cyclisation rate constant
$k_{c}$, according to the reaction equation, 
\[
P_{3}\xrightarrow{k_{c}}P_{3}^{c},
\]
where $P_{3}$ denotes a trimer, and $P_{3}^{c}$ a trimer with a
cycle. Finally, the third line of \eqref{eq:P} refers to appearance
of a loop (in graph theoretical sense), which represents a reaction
firing inside a monomer reducing its functionality by one. The sampling
probability \eqref{eq:P} is proportional to the combinatorial number
of possibilities to select a pair of bonds out of the available free
functionalities, leading to the binomial coefficient $\binom{d_{i,free}}{2}.$
The propensity constant $c_{m}$ is directly equal to reaction rate
constant $k_{m}$ in 
\[
M_{i}\xrightarrow{k_{m}}M_{i-1},
\]
where $M_{i}$ denotes an $i$-functional monomer.

\paragraph{Gillespie's algorithm}

The basis of Gillespie's algorithm is formed by the concept of the
homogeneous Poisson process, that counts reaction events occuring at rate $\lambda$. In this process, the number
of events occurring in a time interval $(t,t+\tau]$ follows a Poisson
distribution with an associated parameter $\lambda\tau$, 
\[
\mathcal{P}(k)=\frac{e^{-\lambda\tau}(\lambda\tau)^{k}}{k!}.
\]
Thus, the waiting times between events are distributed according to
\begin{equation}
p(\tau)=\mathcal{P}(0)=e^{-\lambda\tau}.\label{eq:time}
\end{equation}
Gillespie proposed a stochastic simulation algorithm (SSA), where
an exponential distribution \eqref{eq:time} is used to estimate the time
intervals between reaction firings \cite{gillespie2007}. In the SSA
the event rate is not constant but depends on the reaction propensity
$\lambda_{z},$ (in the current case introduced in \eqref{eq:lambda}).
The actual time between the start and stage $z$ is a sum of $z$
waiting times: 
\begin{equation}
t(z)=-\sum _{i=1}^{z}\frac{\ln(u_{i,0})}{\lambda_{i}},\;u_{i,0}\sim U[0,1]\label{eq:wait.times}
\end{equation}
where $u\sim U[0,1]$ refers to a uniformly distributed random variable.
Therefore, the reaction time $t(z)$ is a random variable itself,
that is determined by $z$ samples from a uniform distribution.

On every stage of the reaction $z,$ a pair of nodes $i,j$ is selected
to receive an edge according to the probability distribution $P(\boldsymbol{A})_{i,j},$
given in \eqref{eq:P}. Each of these selections is uniquely defined
by two uniformly distributed random variables, $u_{z,1},u_{z,2}\sim U[0,1].$
This implies that the adjacency matrix at stage $z$ is uniquely defined
by a specific sampling history. More precisely, the adjacency matrix
$A$ on stage $z$ can be put into a bijective correspondence to a
$z\times2$ matrix that defines samples on the preceding stages: 
\begin{equation}
\boldsymbol{A}(z)=F(\boldsymbol{U}),\boldsymbol{U}=\left(\begin{array}{c}
u_{0,1},\;\;\;\;u_{0,2}\\
u_{1,1},\;\;\;\;u_{1,2}\\
\dots\\
u_{z-1,1},u_{z-1,2}
\end{array}\right).\label{eq:U}
\end{equation}
This reflects the usual idea behind Monte Carlo simulations that all
networks $F(U)$ for all sample matrices $\boldsymbol{U}$ are considered
as equiprobable. Consequently, any statistical property (\latin{e.g.} average
degree, component size distribution, density) of the network $\phi(\boldsymbol{A})$
may be determined by an integration over the set of sampling variables
$\Omega=[0,1]^{2t}$. If independent sampling matrices $\boldsymbol{U}$
were drawn $N$ times, the integral is approximated by a sum, the
Monte Carlo estimator: 
\begin{equation}
\int\limits _{\Omega}\phi(F(\boldsymbol{U}))du\approx\frac{1}{N}\sum _{i=1}^{N}\phi(F(\boldsymbol{u}_{i})).\label{eq:est}
\end{equation}
Here, $\boldsymbol{U}_{i},\;i=1,\dots,N$ are randomly generated instances
of matrix \eqref{eq:U}. The error of estimator \eqref{eq:est} decreases
asymptotically as $N^{-\frac{1}{2}}$ when $N\rightarrow\infty.$

\paragraph{Markov Chain Monte Carlo, simulation of a large system}

The stochastic simulation that has been described above, proceeds
by drawing samples distributed according to the two-dimensional probability
distribution $P(\boldsymbol{A})_{i,j}$ given in \eqref{eq:P}. This
can be done by drawing two successive samples from one-dimensional
distributions. The first sample, $s_{1},$ is drawn from the marginal
distribution $\sum _{i=1}^{n}P(\boldsymbol{A})_{i,j}$; the
second, $s_{2}$ - from the conditional probability distribution $P(\boldsymbol{A})_{s_{1},j}\Big(\sum _{j=1}^{n}P(\boldsymbol{A})_{s_{1},j}\Big)^{-1}.$
This procedure requires recomputing all entries of the matrix $P(\boldsymbol{A})_{i,j}$
, which is computationally the most expensive step of the whole algorithm.
Observe, that even though the adjacency matrix $\boldsymbol{A}$ is
very sparse, matrix $P(\boldsymbol{A})_{i,j}$ with its upper bound
on density $E_{max}/n^{2},$ is a full one.

The computational cost of the algorithm may be significantly reduced
utilising a Markov Chain Monte Carlo (MCMC) algorithm, which is also
commonly used in statistical mechanics and other disciplines. The
MCMC algorithm replaces the draws by a specific Markov process having
$P(\boldsymbol{A})_{i,j}$ as the stationary distribution \cite{brooks2011}.
More specifically, for a fixed matrix $\boldsymbol{A}(z)$, we design
a sequence of pairs $\big(s_{1}(q),s_{2}(q)\big)$ that is initiated
by two random numbers from $1\,\dots,n.$ The transition from pair
$q$ to $q+1$ is realised according to the following principle: 
\begin{equation}
(s_{1}(q+1),s_{2}(q+1))=\begin{cases}
\big(s_{1}^{*}(q),s_{2}^{*}(q)\big), & \text{if}\;u(q)<\min\Big(\frac{P_{s_{1}^{*}(q),s_{2}^{*}(q)}}{P_{s_{1}(q),s_{2}(q)}}\frac{P_{s_{1}(q),s_{2}(q)}^{*}}{P_{s_{1}^{*}(q),s_{2}^{*}(q)}^{*}},1\Big)\\
\big(s_{1}(q),s_{2}(q)\big), & \text{otherwise}
\end{cases}\label{eq:MCMC}
\end{equation}
where $u(q)\sim U[0,1]$, and $(s_{1}^{*}(q),s_{2}^{*}(q))$ is drawn
from the proposal distribution $P^{*}(\boldsymbol{A})$. The proposal
distribution coincides with the original $P^{*}(\boldsymbol{A})_{i,j}=P(\boldsymbol{A})_{i,j}$
when $i=j,$ and assumes a simpler form for $i\neq j,$ 
\begin{equation}
P^{*}(\boldsymbol{A})_{i,j}=\lambda^{*}\cdot\left\{ \begin{aligned} & \,c_{e}\,d_{i,free}d_{j,free}\Big(g_{i}(\boldsymbol{A})g_{j}(\boldsymbol{A})\Big)^{\beta} & \text{ intermolecular };\\
 & \,c_{i}\,d_{i,free}d_{j,free}, & \text{intramolecular},i\neq j;\\
 & \,c_{m}\,\binom{d_{i,free}}{2}, & i=j.
\end{aligned}
\right.\label{eq:Pstar}
\end{equation}
Although the MCMC scheme \eqref{eq:MCMC} provides only a means of
approximate sampling, it possesses one very important feature: it
is not necessary to compute all the elements of matrix $P_{i,j}$
but only certain selected values. This feature allows to considerably save on computational
resources for large adjacency matrices.

The downside of MCMC is that one has to ensure that the Markov chain
\eqref{eq:MCMC} operates in a stationary regime before terminating
it and accepting the final values as a random sample that mimics samples
from $P_{i,j}$.

\section{Discussion/application to linseed oil network}

In this section we discuss the kind of information one may extract
from the random molecular networks generated by the stochastic simulations
using the Random Graph Model as introduced above. The results are
formulated as graph-theoretical properties that may be calculated
with the MC estimator \eqref{eq:est}. These data are useful in improving
our understanding of the system, but also might be used for physical
interpretation of the topological information and its influence on
the observable and measurable properties of the final material.

\begin{figure}[h]
\center \includegraphics[width=0.9\textwidth]{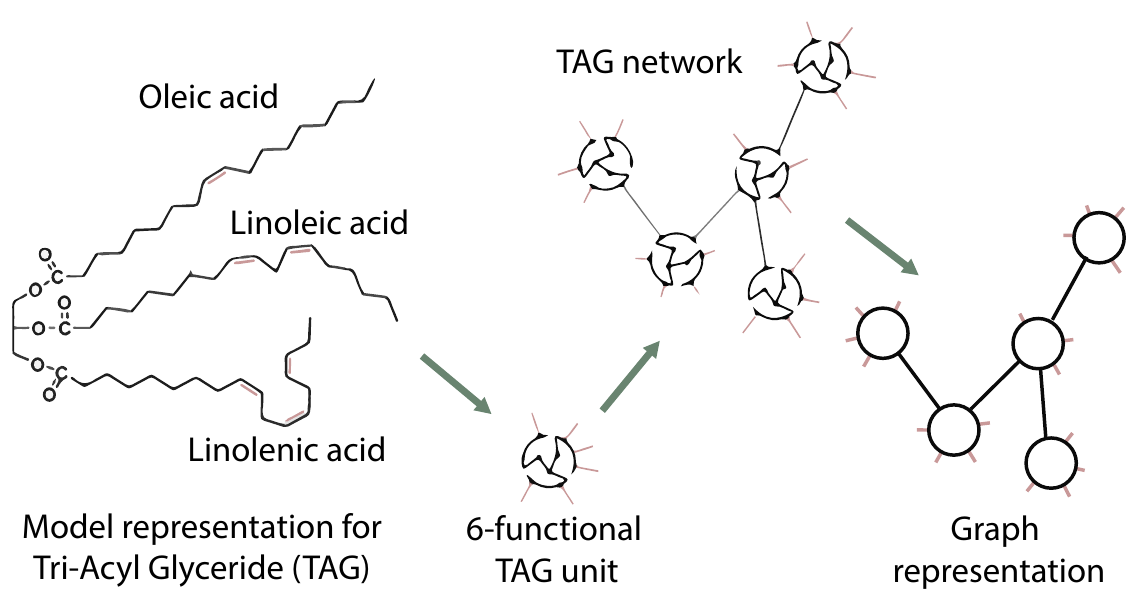} \caption{ Conceptual model for representation of the linseed oil network. Try-acyl
glyceride is treated as a single node. Depending on composition of
the nodes, they may have maximum functionality ranging between 0 and
9. }
\label{fig:TAG} 
\end{figure}

\begin{figure}
\center \includegraphics[width=0.65\textwidth]{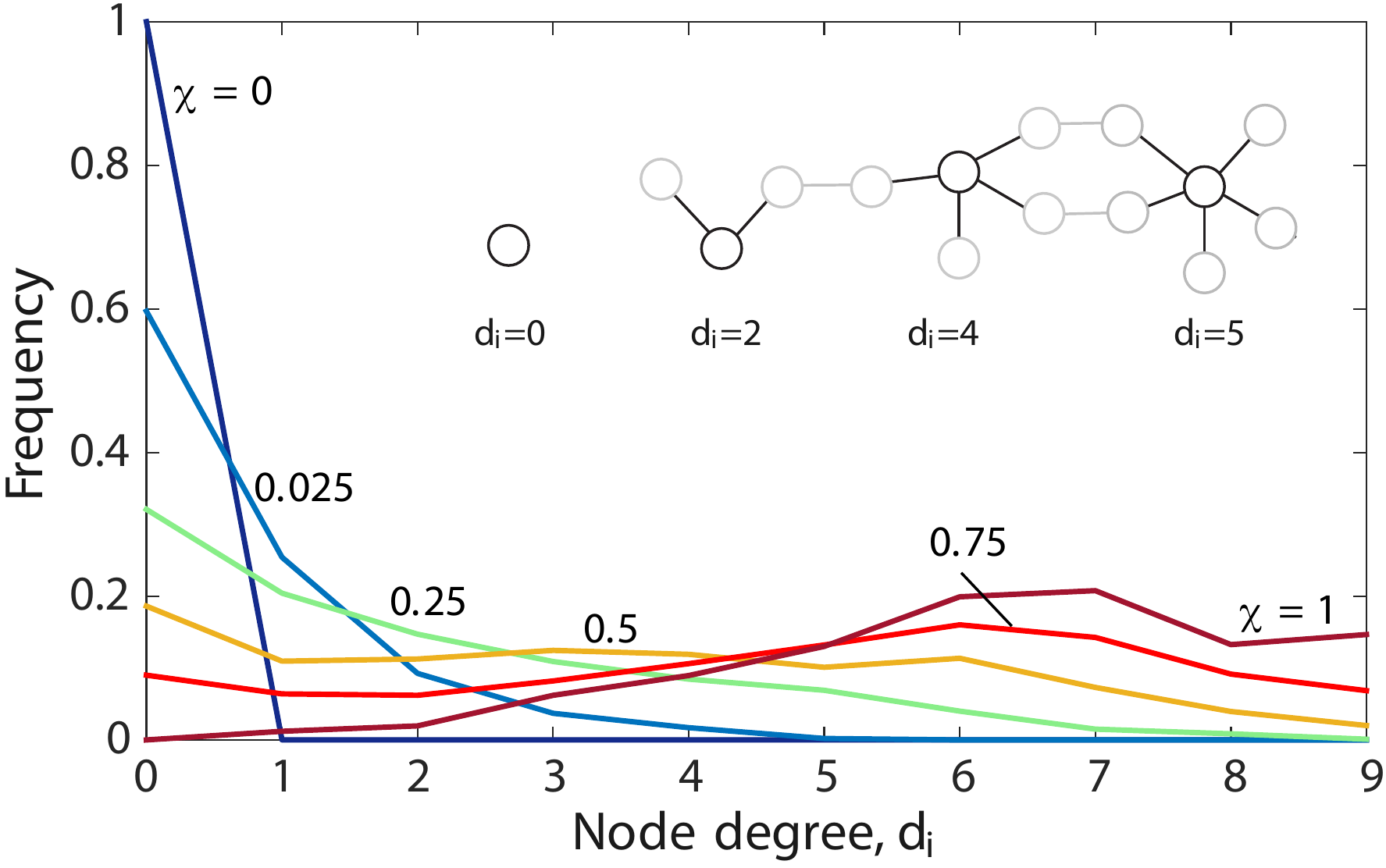} \caption{ Evolution of degree distribution during polymerisation. Fraction
of edges, $\chi\in[0,1],$ depicts the progress of the stochastic
process. The final distribution at $\chi=1$ coincides with distribution
of TAG-unit functionalities in linseed oil.}
\label{fig:degrees} 
\end{figure}

\begin{figure}
\center \includegraphics[width=0.65\textwidth]{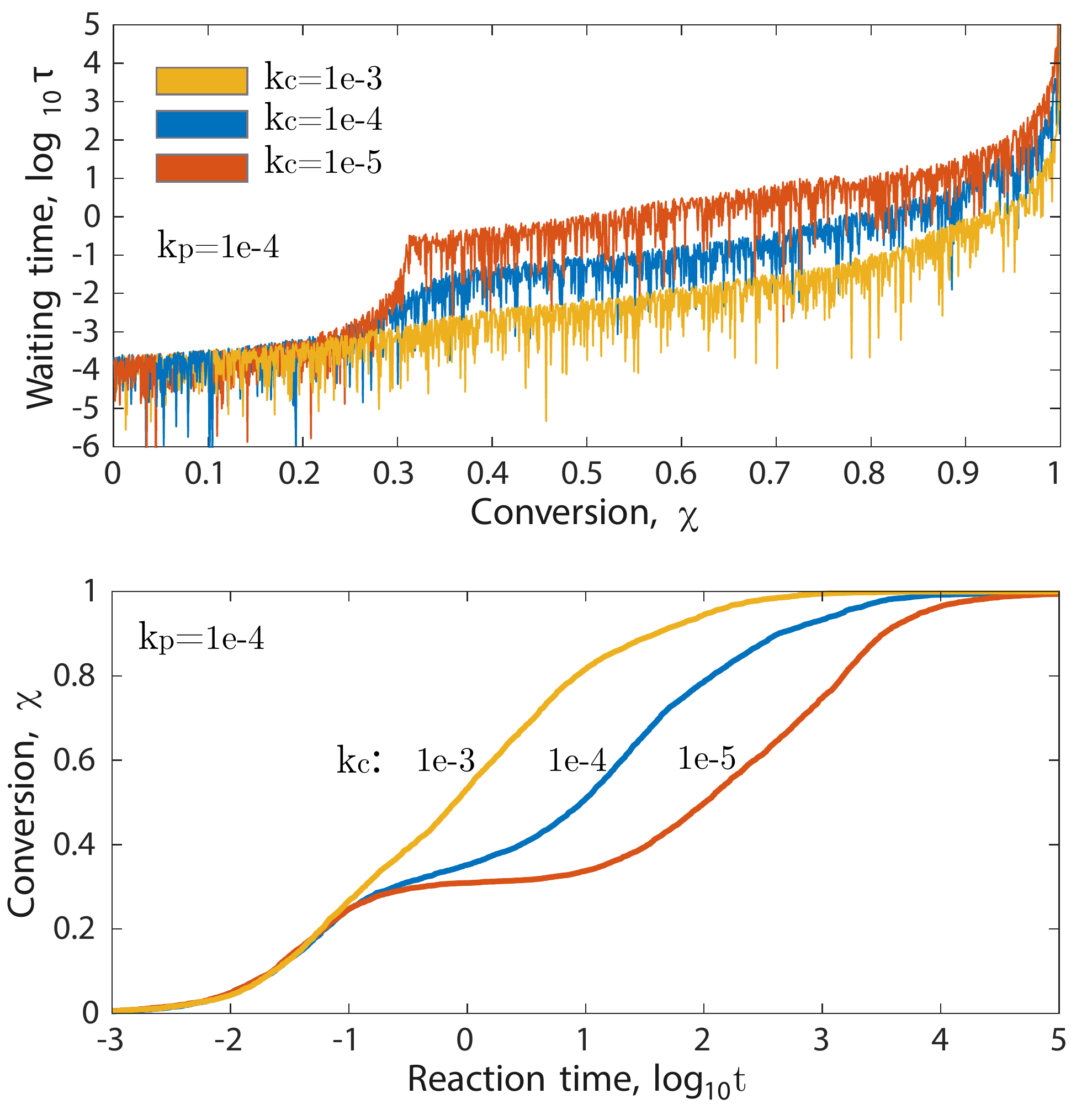} \caption{ (\emph{Top:}) time span between reaction events as a function of
the conversion. (\emph{Bottom:}) the conversion of edges versus total reaction
time. }
\label{fig:times} 
\end{figure}

In order to demonstrate the RGM approach on a practical case we consider
linseed oil polymerisation, which is more commonly known as the 'drying'
of oil paint. The binding medium linseed oil consists of triacylglycerides
(TAG) and the unsaturated groups on the three fatty acid (C19) chains
cause the TAG-units to polymerise into a network. This curing process
is a complex process involving radicals, oxygen, peroxides, etc.,
which is only partly understood, and only little quantitative kinetic
information is available \cite{iedema2014Mel}. Here, we simplify this
complex process to the coupling of the TAG-unit star monomers according
to an estimated 'chemical' coupling rates, $k_{c}$. and $k_{p}$,
which then is affected by decreasing functionality, distance within
the network and steric hindrance as is now possible with the RGM model.
We expect that this model with strongly simplified kinetics will nevertheless
contribute to a basic understanding of the linseed oil network, as
influenced by its curing behaviour, but also by degradation, for instance
by hydrolysis of the ester bonds. The conceptual model of TAG-unit
network is illustrated in Figure~\ref{fig:TAG}. TAGs contain double
bonds that are responsible for the formation of the cross-links. Since
there can not be more then 3 unsaturations per fatty acid in the TAG
node, the maximum number of connections node $i$ can form is limited
to 9, $0\leq d_{i,max}\leq9$. More precisely $d_{i,max}$ are distributed
according to frequency distribution $W_{n}=\sum _{i+j+k=n}f_{i}f_{j}f_{k},\;n=0,...,9,$
where $f_{i}$ denotes concentration of fatty acids with $i$ unsaturations.
The frequency that we have assumed in this study is based on the natural
abundance of oleic, linoleic and linolenic acid in linseed oil, having,
respectively, one, two and three unsaturations (see Figure 5). From
the point of view of a single node the polymerisation process looks
simple: the degree of each node, $d_{i}$, will transit from 0 to,
eventually $d_{i,max}$. Thus the evolution of the frequency distribution
of node degrees, 
\[
f_{i,degree}(\boldsymbol{A})=\frac{1}{n}\sum _{j=1}^{n}\delta(d_{j}(A)-i).
\]
has a simple form: in the beginning of the polymerisation process
the distribution describes disconnected monomers, $f_{i,degree}(\boldsymbol{A}(0))=0,i>0,$
and $f_{0,degree}(\boldsymbol{A}(0))=1.$ At the full conversion the distribution
of node degrees approaches the distribution of functionalities of
pure monomers (Figure~\ref{fig:degrees}).

The global properties of the network have less trivial dynamics. In order to extract various network properties we simulated ensembles of $10^4$ monomers and averaged the results over 100 of such ensembles. These
global properties in turn affect reactivity rates and the overall
kinetics of the system. Even the timing of reaction events, given
in \eqref{eq:wait.times}, strongly depends on the ratio of rate constants
$\nicefrac{k_{c}}{k_{p}}.$ For time intervals between the reaction
events and the overall conversion/reaction time plot see panels in
Figure~\ref{fig:times}. Alternatively, instead of increasing $\nicefrac{k_{c}}{k_{p}},$
one may consider to decrease concentration of monomers, which by
enhancing cyclisation has an identical effect on the system.

\paragraph{Gel transition}

\begin{figure}
\center \includegraphics[width=0.7\textwidth]{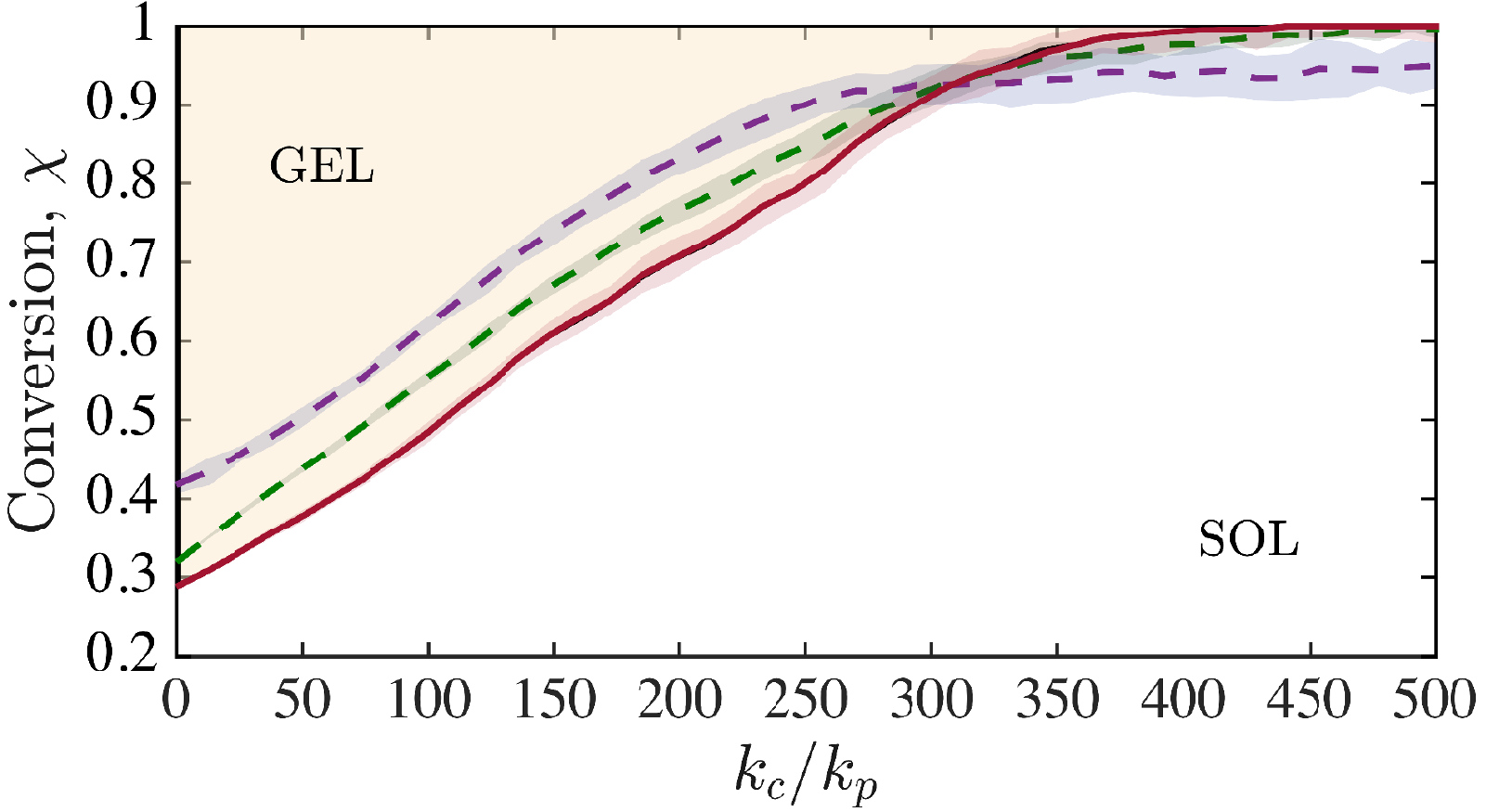} \caption{ Gel point conversion as a function of reactivity ratio as obtained
from the RGM simulation, confidence interval for $p=0.95$. The dashed lines indicate the point when the average shortest path length is maximum. }
\label{fig:gelpoint} 
\end{figure}

Polymerisation of multifunctional monomers is known to pass through
the gel transition: a transition that drastically affects the physical
properties of the material. The gel point is experimentally well observable
due to considerable differences in the material properties between
the gel and sol phases, such as solubility and viscosity \cite{bower2002}.
Hence, it is a common practice to define the gel transition point as
a point where the observable properties change. Here, we will define
the gel transition entirely on topological properties of the corresponding
graph. Let's consider a sequence of adjacency matrices $A(n)$ constructed
for identical input parameters and given number of nodes, $n$. We
associate the gel part of the network with the giant component (in
the sense of random graph theory \cite{spencer2010}). Thus, the gel
point is the time instant in the reaction progress at which the giant
component emerges. More specifically, the largest component in the
network is called gel if its size $S_{1}(n)=O(n)$ when $n$ approaches
infinity. Furthermore, if gel is present, the size of the second largest
component is $S_{2}(n)=O(\log(n))$. In the pre-gel regime both $S_{1}(n)$ and $S_{2}(n)$
are equal to $O(\log(n)).$ Thus distinguishing between pre- and post-gel
regimes is equivalent to distinguishing between two asymptotical modes
of $S_{2}(n)/S_{1}(n)$ when $n\rightarrow\infty.$ We use this asymptotic
estimates to determine whether the system operates in the gel regime
for a given progress of the reaction $z$ (or conversion $\chi$),
\[
g_{01}(\chi)=\begin{cases}
0, & \text{if }\;S_{2}(n)/S_{1}(n)\rightarrow const\\
1, & \text{if }\;S_{2}(n)/S_{1}(n)\rightarrow0
\end{cases}
\]
where $S_{1}(n)$ is the size of the largest component for a system
with $n$ monomers and $z$ edges, and $S_{2}(n)$ - size of second
second-largest component. Clearly, $g_{01}(\chi)$ itself is a random
variable and its mean has to be estimated by the MC estimator. When
the number of samples in the MC estimator $N$ becomes large, the
gel indicator function $g_{01}(\chi)$ approaches the Heaviside function
$H(\chi-\chi_{g}),$ where $\chi_{g}$ is a 'breaking point' separating
the pre-gel and gel regimes. In practice, for a fixed $N$ , we identify
$\chi_{g}$ by solving a least-squares problem that minimises the
residual $\int\limits _{\chi}|H(\chi-\chi_{g})-g_{01}(\chi)|d\chi.$

The gel point conversion $\chi_{g}$ is plotted versus $\nicefrac{k_{c}}{k_{p}}$
in Figure~\ref{fig:gelpoint}. A clear discontinuity can be seen:
the fraction equals to 0 up to the gel transition. Higher levels of
$\nicefrac{k_{c}}{k_{p}}$ postpone the gel transitions up to the
point when gelation does not occur at all. This is a logical result,
since a higher $k_{c}$ implies a stronger intramolecular cross-linking,
which obviously competes the gel formation.

\paragraph{Molecular sizes}

\begin{figure}
\center \includegraphics[width=0.8\textwidth]{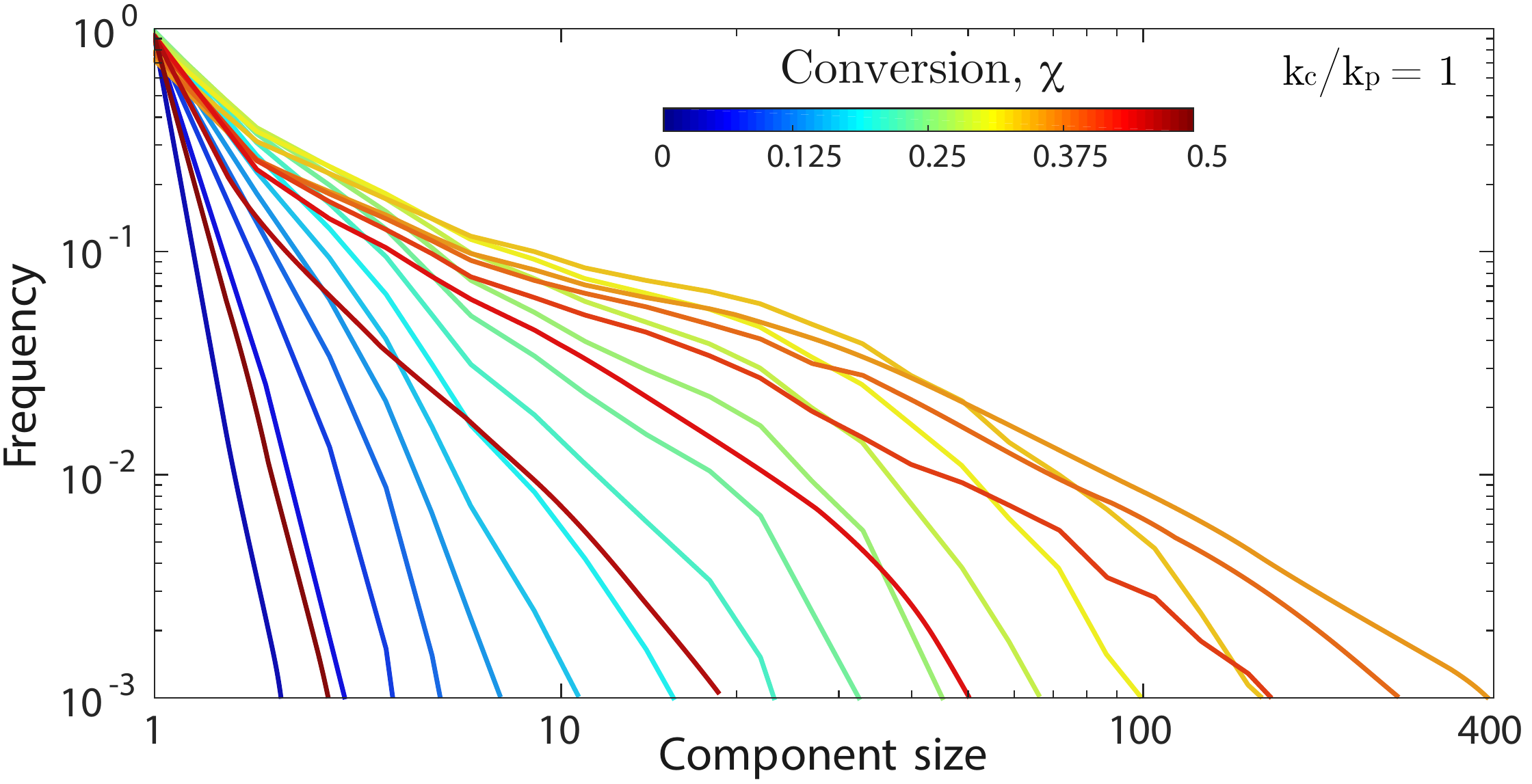} \includegraphics[width=0.8\textwidth]{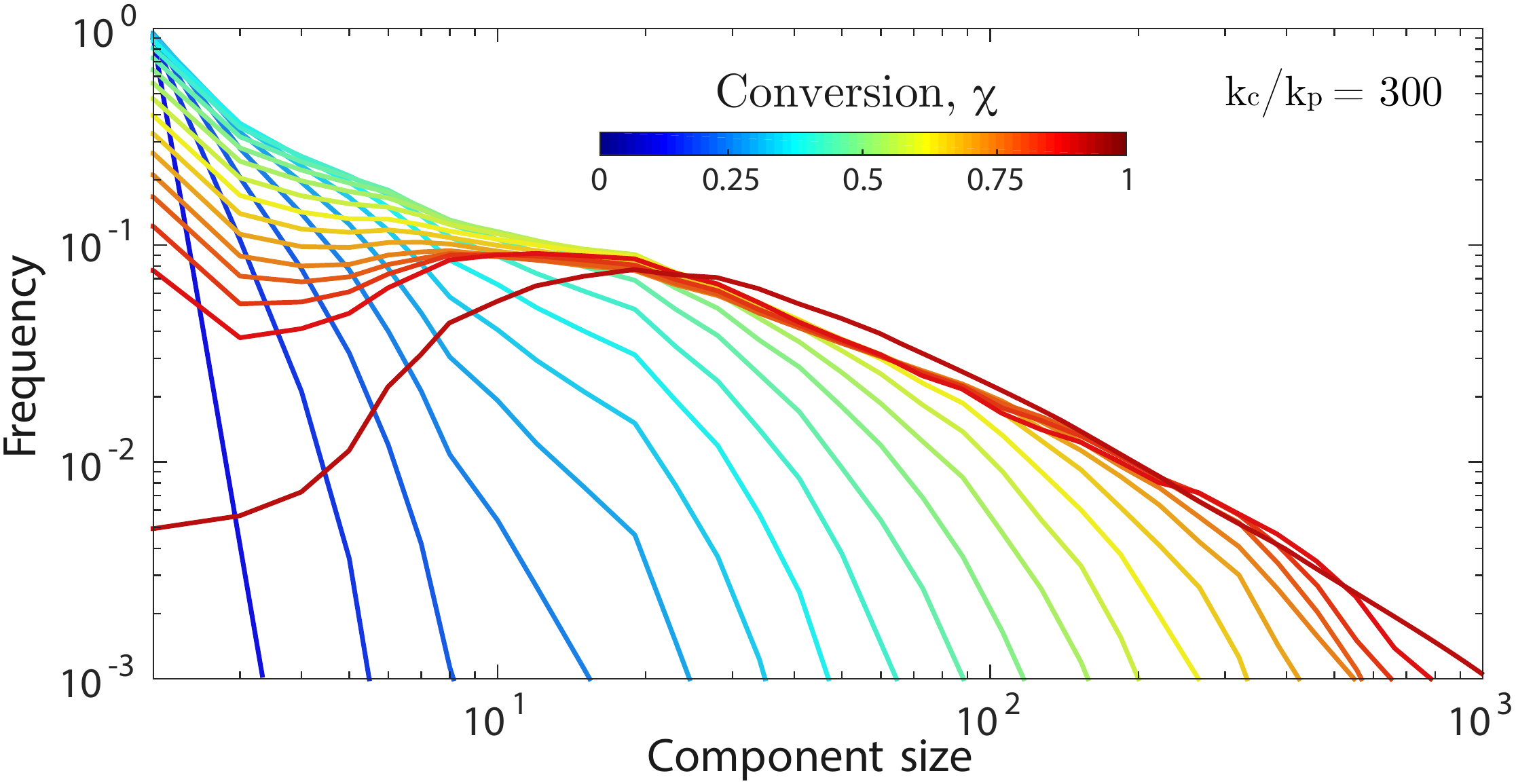}
\includegraphics[width=0.8\textwidth]{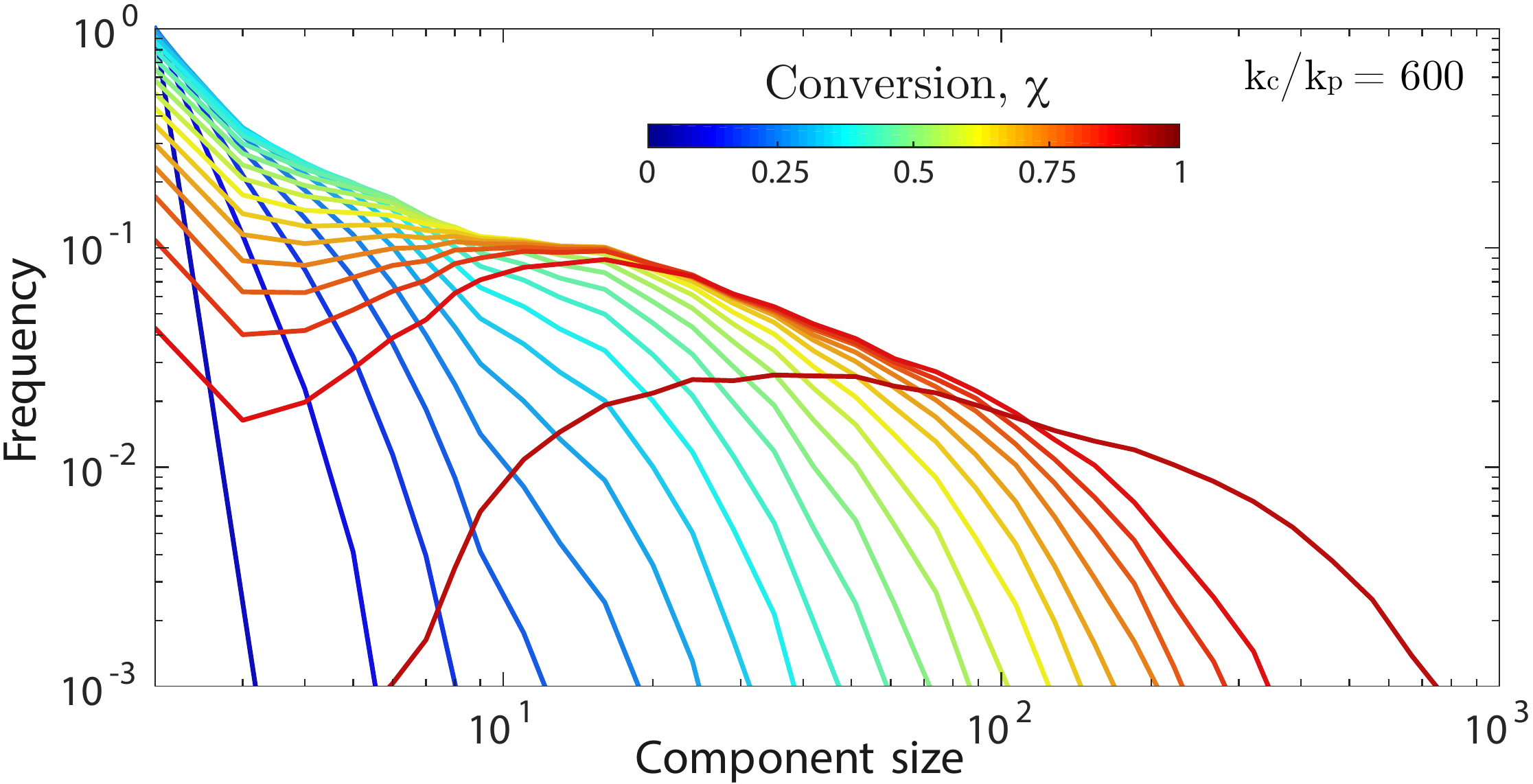}

\caption{Evolution of size distributions (excluding gel) for three values of
$\nicefrac{k_{c}}{k_{p}}$. }
\label{fig:sizes} 
\end{figure}

As shown in Figure~\ref{fig:times} the reaction rates are not constant,
but slow down in a non-linear fashion. One of the causes for this
effect is the emergence of large, connected components that restrict
their internal monomers in reactivity. Molecular sizes in terms of
number of monomers are equal to the number of nodes in connected components
of the network that are not connected to each other. It is straightforward
to identify the connected components of a graph in linear time using,
for instance, breadth-first search\cite{jungnickel2008}. Figure~\ref{fig:sizes}
shows how the distribution of sizes of these components (molecular
size distribution) evolves in time. 

\begin{figure}
\center \includegraphics[width=0.7\textwidth]{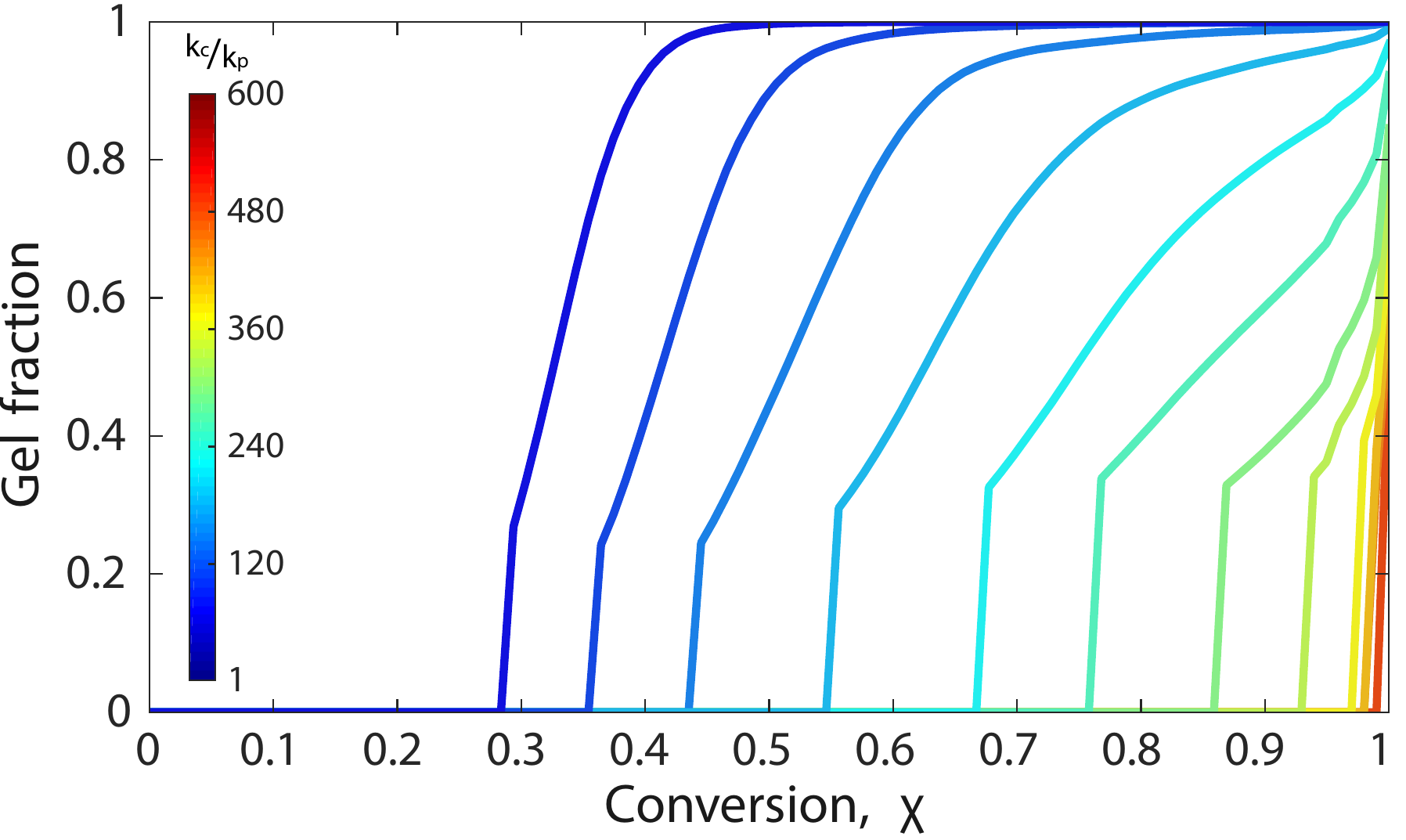} \caption{ Fraction of monomers involved in giant component (gel) as a function
of conversion for various values of $\nicefrac{k_{c}}{k_{p}}$. }
\label{fig:gel.fraction} 
\end{figure}

Eventually, a connected component of the same order of magnitude as
the whole network emerges (the giant component). The point in time/conversion
when this event takes place marks the usual gel transition. The size
of the giant component increases rapidly as it becomes connected to
the rest of the components. Reactions between large components (including
the giant component) dominate over reactions between small-sized components.
This is why the size distributions in Figure~\ref{fig:sizes} shifts
back towards the origin after the gel transition. Thus, one observes
that the gel point detected from the topology alone, as explained
above, indeed marks the turning point of the size distribution. For
high values of $\nicefrac{k_{c}}{k_{p}}$ the gel transition is postponed
up to the full conversion (Figure~\ref{fig:sizes}b) or does not happen
at all (Figure~\ref{fig:sizes}c). According to the general convention
the data presented in Figure~\ref{fig:sizes} excludes the giant
component itself as it only refer to the sol part. In a complementary
fashion, the fraction of nodes incorporated in the giant component
is given in Figure~\ref{fig:gel.fraction}. 

\paragraph{Structure of the network}

The characterisation of the polymer network in terms of numbers and
sizes of connected components is a widely used in the polymer community.
However, in cases where the topology is not a priori known (\latin{e.g.} linear
or regularly branched polymers), we have to invent new characteristic
measures to describe how individual monomers are arranged inside the
single connected component that the network eventually becomes. Although
being randomly interconnected, the nodes may occasionally line up
into certain \emph{motifs} of special interest. The search for motifs
that are frequently occurring in the network is a new structural characterisation
method that is easy to quantify. For instance, the degree distribution
is nothing else than the frequency of nodes having zero, one, two,
etc., adjacent nodes.

\begin{figure}
\center \includegraphics[width=0.7\textwidth]{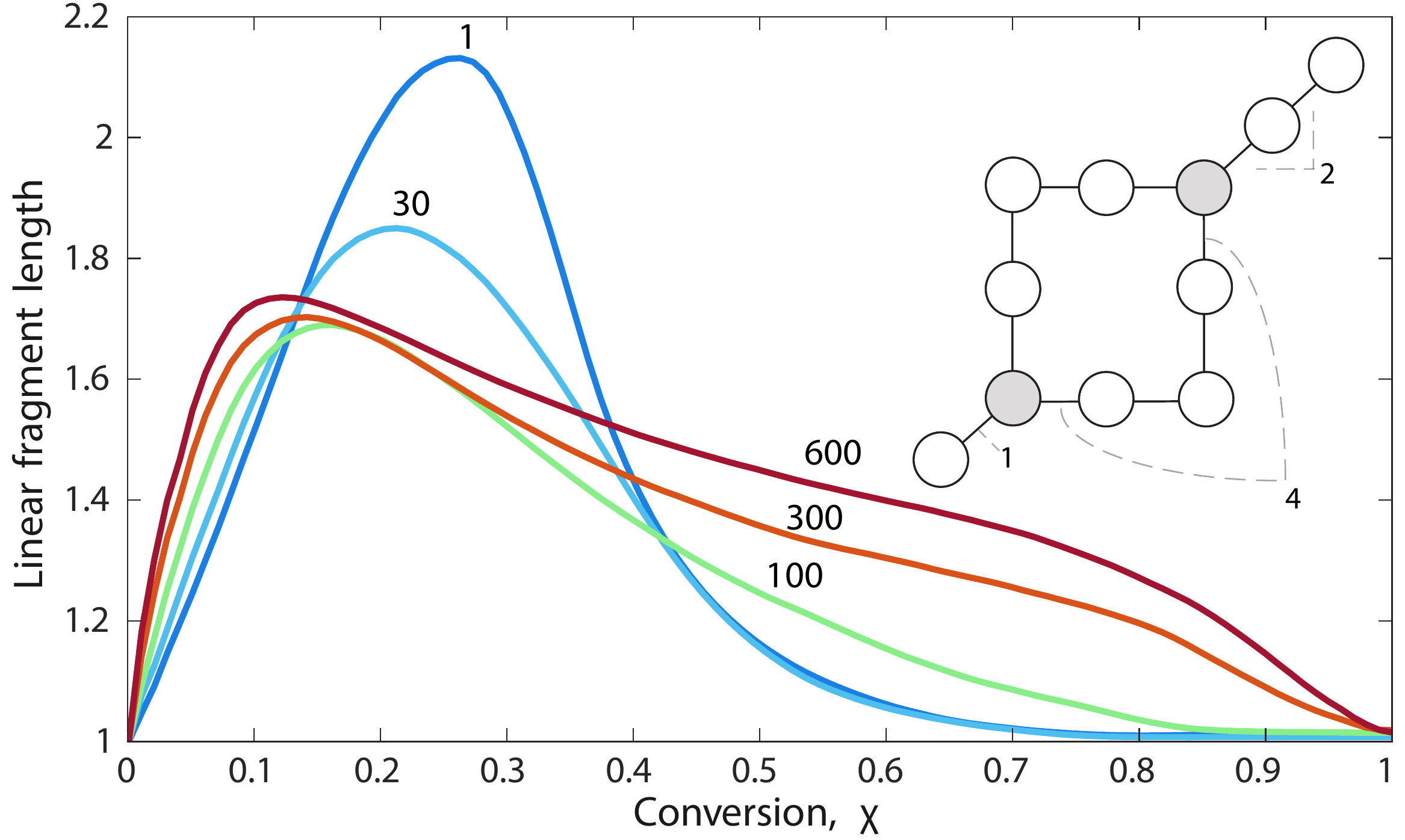} \caption{ Average length of linear fragments as a function of conversion for
various values of $\nicefrac{k_{c}}{k_{p}}.$ The illustration depicts
a sample network with four linear fragments of lengths: 1, 2, 4, 4. }
\label{fig:linear} 
\end{figure}

Besides structural information, the size distribution of linear fragments
plays an important role in the elasticity of polymeric networks \cite{gunduz2008,ngai1994}.
The number-average length of linear fragments strongly depends on
the conversion; it increases at early stages of the reaction and eventually
decreases to almost to 1. High values of $\nicefrac{k_{c}}{k_{p}}$
are associated with shorter linear fragments at intermediate stages
of conversion, but ultimately, with longer linear fragments at later
stages (Figure~\ref{fig:linear}). 

Another motif-related property is the local clustering coefficient.
For node $k$, it is equal to fraction of pairs connected to $k$
that are also connected to each other(\latin{i.e.} forms a triadic clousure), 
\[
c_{k}(\boldsymbol{A})=\frac{\sum _{i<j}\boldsymbol{A}_{i,j}\boldsymbol{A}_{i,k}\boldsymbol{A}_{j,k}}{d_{k}(d_{k}-1)}.
\]
This coefficient is a number between 0 and 1, see the example of Figure
12. In order to transit from local clustering per node to average
clustering of the whole network we apply averaging $\boldsymbol{A},$
$c(\boldsymbol{A})=\frac{1}{n}\sum c_{k}(\boldsymbol{A})$. 
\begin{figure}
\center \includegraphics[width=0.7\textwidth]{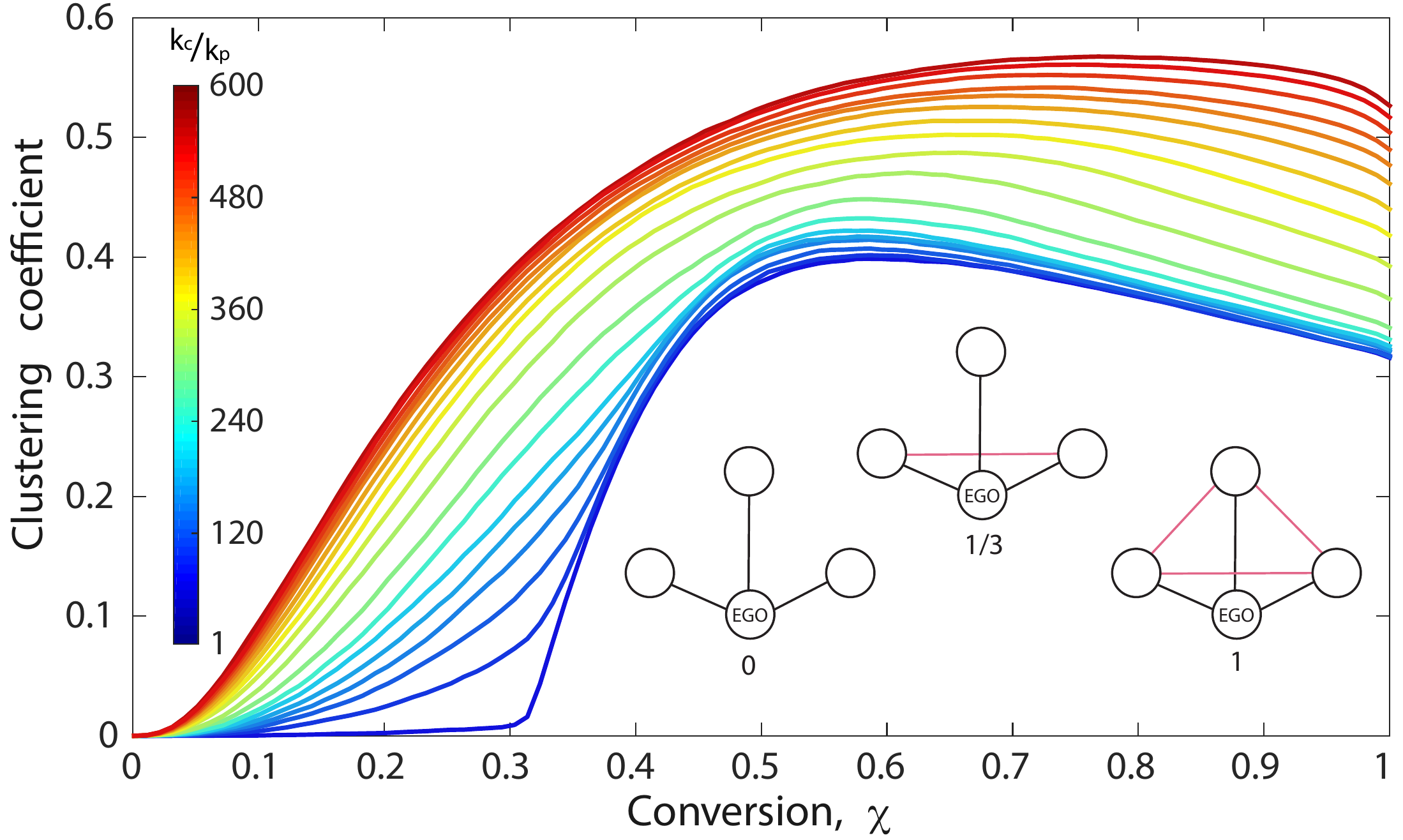} \caption{Clustering coefficient as a function of conversion for various values
of $\nicefrac{k_{c}}{k_{p}}$. Three examples of a 4-node network
are given. The clustering coefficient of the node marked 'EGO' is
equal to 0, 1/3, or 1 depending on the connections between EGO's neighbouring
nodes. }
\label{fig:clustcoeff} 
\end{figure}

The average clustering coefficient gives a good idea of small scale
patterns that may arise in the network. At the beginning of the polymerisation,
the average clustering increases monotonically in time. For systems with
low values of $\nicefrac{k_{c}}{k_{p}}$ the average clustering becomes
significant only in the gel regime. It is interesting to note that
the average clustering decreases at conversions close to 1. This effect
is caused by steric hindrance in the ultimately overcrowded network
that eventually suppresses clustering.

\begin{figure}
\center \includegraphics[width=0.98\textwidth]{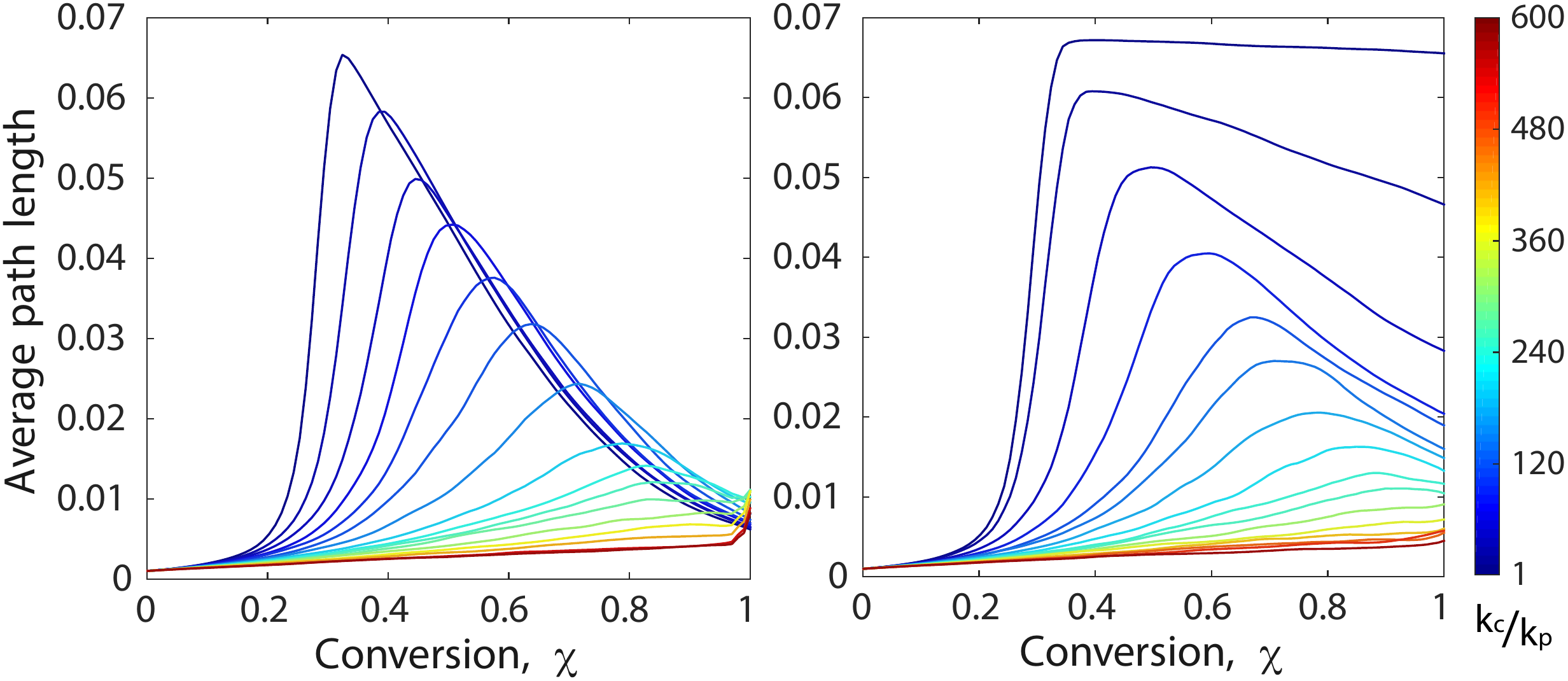} \caption{ Average path length as a function of conversion. Left panel depicts
results in the case of no intramonomeric reactions, right panel -
high intramonomeric reaction rate ($\nicefrac{k_{i}}{k_{p}}=10$). }
\label{fig:path.length} 
\end{figure}

\paragraph{Average path length}

The average path length of a network is a concept that defines the
average number of steps along the shortest paths for all possible
pairs of network nodes, 
\[
L(\boldsymbol{A})=\frac{1}{n(n-1)}\sum_{i\ne j}p_{i,j},
\]
Here, $p_{i,j}$ denotes the distance between nodes $i$ and $j$
in network $\boldsymbol{A}$, if the nodes $i,j$ are not connected,
$p_{i,j}=0.$ One may think of average path length as a measure of
'network density'. Various theories connect the shortest path to observable
data related to light scattering, radius of gyration, etc. \cite{rouvray2002}.
Figure~\ref{fig:path.length} shows how the normalised average path
length evolves during reaction progress for various values of $\nicefrac{k_{c}}{k_{p}}.$
Note that even though the average path length is normalised by dividing
by the total number of nodes $n$, it still depends on $n$ and special
care has to be taken, when comparing the average path lengths of two
systems of different size. The two panels presented in Figure~\ref{fig:path.length}
represent cases with and without intramonomeric reaction. A pattern
that is typical for many polymer properties, can be seen: the average
path increases in the pre-gel regime and decreases thereafter. However,
closer investigation reveals that the maximum of the average path
does not exactly coincide with the gel transition but is shifted.
This is demonstrated by Figure~\ref{fig:gelpoint}, where the conversion
at maximum average path length and at the gel point is plotted for various
values of $\nicefrac{k_{c}}{k_{p}}.$ The intramonomeric reaction leads to a more sparse network
by consuming functional groups, and thus, reducing functionality of the monomers. Since the intramonomeric reaction
is not hindered by the resulting topology, the influence of this reaction
is most visible after the gel transition that is reflected by a suppressed
decrease of the average path length (see Figures~\ref{fig:path.length},\ref{fig:gelpoint}).

\begin{figure}
\center \includegraphics[width=0.8\textwidth]{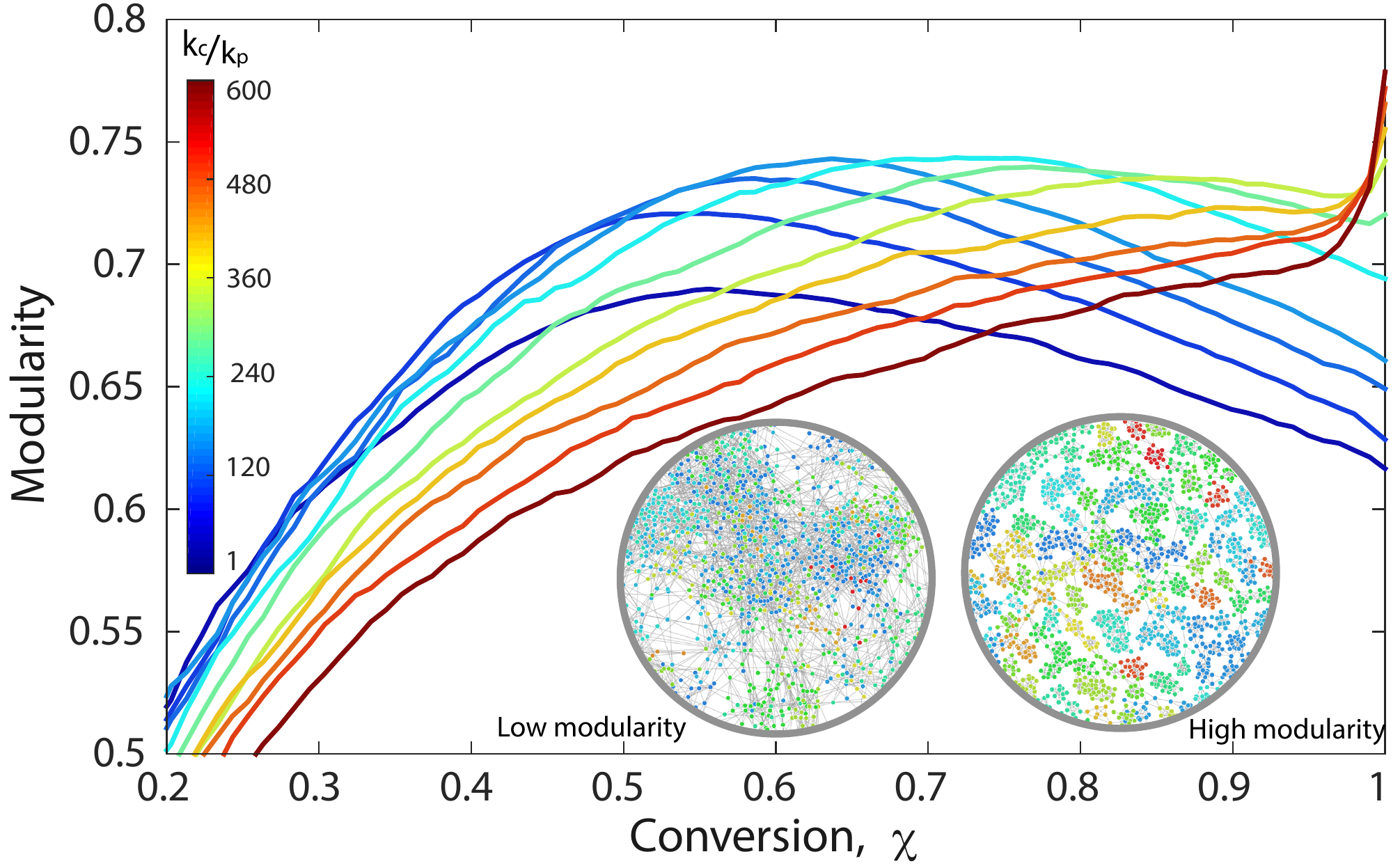} \caption{ Modularity as a function of conversion for various values of $\nicefrac{k_{c}}{k_{p}}.$
The two samples depict networks with low/high modularities. }
\label{fig:modularity.pdf} 
\end{figure}

\begin{figure}
\center \includegraphics[width=0.8\textwidth]{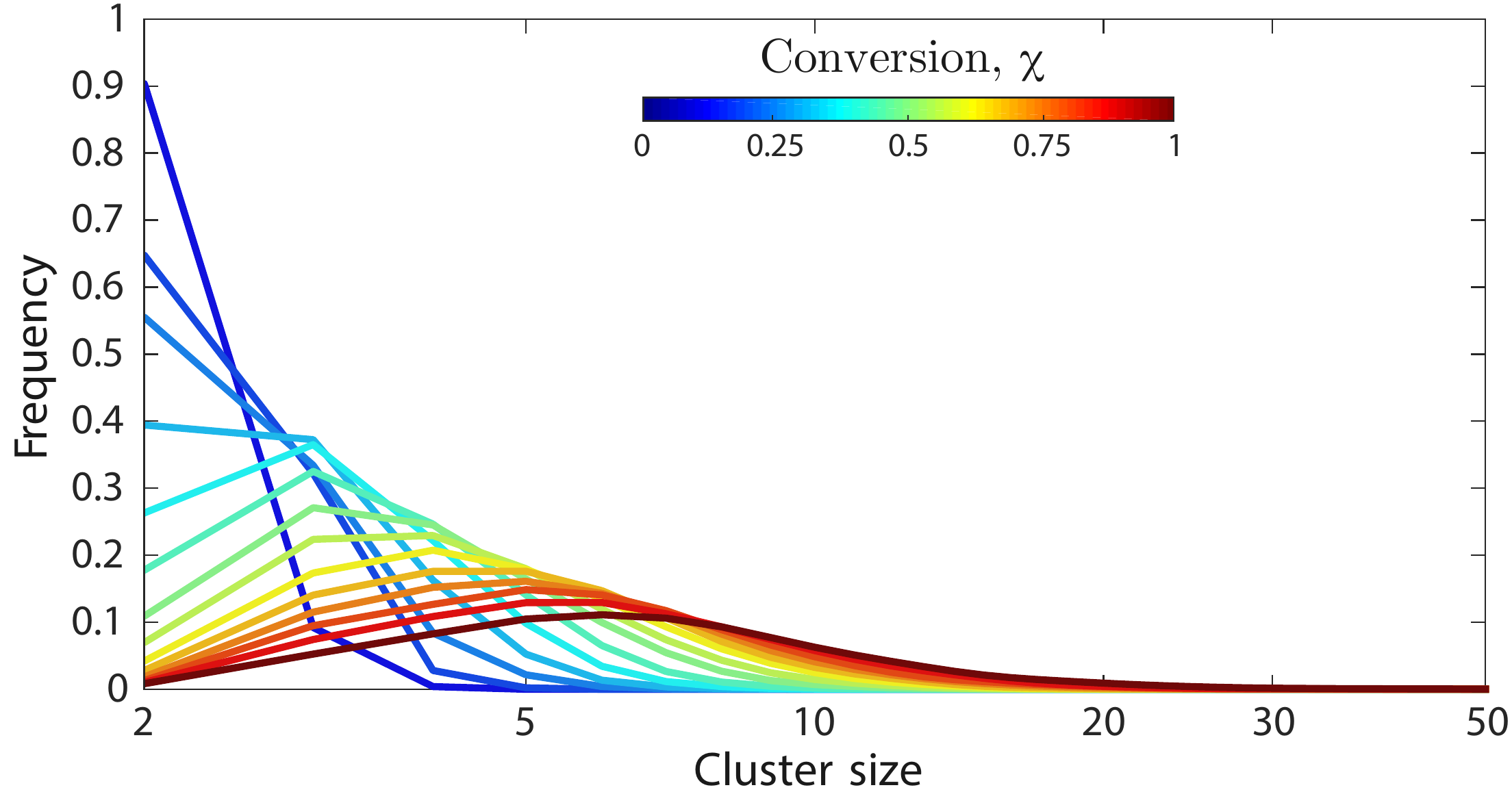}
\caption{ Cluster size distribution for various levels of conversion }
\label{fig:cluster.distribution} 
\end{figure}

\paragraph{Structural non-homogeneity}

The evolving structure of networks may be more or less homogeneous,
for instance microgels may (temporarily) evolve as a form of heterogeneity\cite{seiffert2015,asai2013}.
It turns out that structural non-homogeneity is an inherited property
of the polymerisation models that account for distance between monomers.
Certain values of input parameters lead to topologies that, although
being connected, can be partitioned into distinct clusters, comparable
to microgels. These clusters are loosely connected between each other
but are densely interconnected within themselves. This effect can
be achieved when cyclisation dominates propagation (high values of
$\nicefrac{k_{c}}{k_{p}}$). For example, in diluted systems the chance
that a monomer will connect to its direct neighbour (or another node
from the same component) is higher than the chance of encountering
a disconnected component. One way to measure the degree of inhomogeneity
(clustering) present in the network is the measure of \emph{modularity,}
which may be calculated using a special optimisation algorithm involving
the adjacency matrix \cite{newman2006}. The evolution of modularity
as presented in Figure~\ref{fig:modularity.pdf} exhibits the increase-decrease
pattern with a maximum at intermediate levels of conversion. High
values of $\nicefrac{k_{c}}{k_{p}}$ postpone the maximum towards
the full conversion, up to the point where the modularity increases over
the whole time span of the process. It can be observed that a certain
degree of structural inhomogeneity is prevalent even for very low
values of $\nicefrac{k_{c}}{k_{p}}$ at intermediate conversions.
However, at the final conversion a structure with distinct clusters is
preserved only for high values of $\nicefrac{k_{c}}{k_{p}}.$

As similar to the component size distribution, we may consider a cluster
size distribution that evolves over time with reaction progress (Figure~\ref{fig:cluster.distribution}).
Even though the value of $\nicefrac{k_{c}}{k_{p}}$ defines how well
separated the clusters are, it has little effect on cluster sizes.
However, cluster sizes are strongly affected by excluded volume
constant $\alpha,$ that appears in \eqref{eq:Phi}.

\begin{figure}
\center \includegraphics[width=0.8\textwidth]{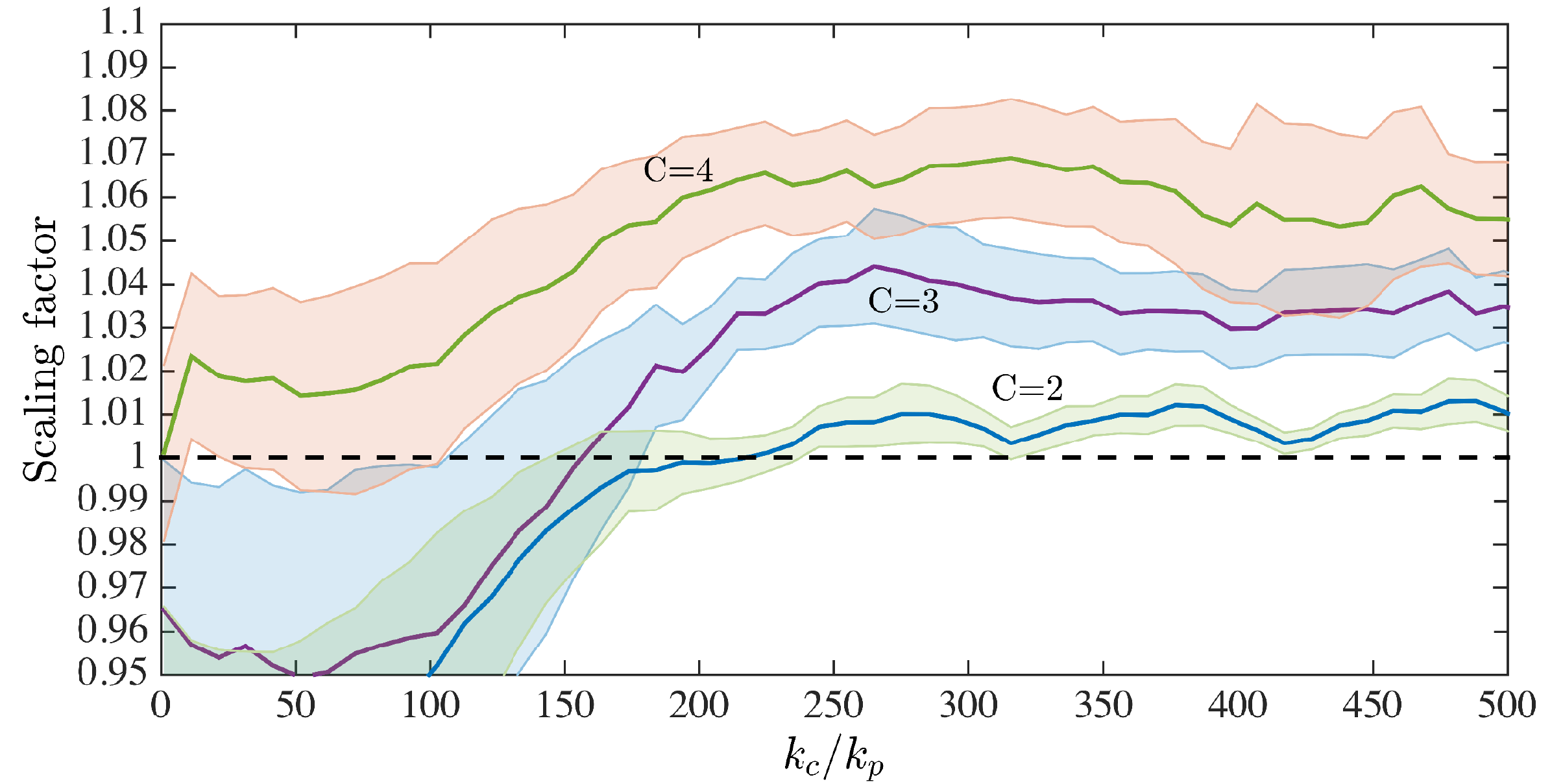} \caption{Scaling fraction vs $k_{c}/k_{p}$ as obtained for three values of
chain stiffness constant, $\alpha$. The bands represent 0.9 coinfidence
interval around mean value based on Monte Carlo sampled data. }
\label{fig:scaling} 
\end{figure}

\paragraph{Scaling}

The average path length is related to the gyration radius of individual
molecules and to the density of the gel in space. The scaling of the
average path length for gel networks has to satisfy the inequality
\eqref{eq:scaling}. Testing this inequality for various values of
input parameters helps to narrow down the range of valid values for
these parameters. Figure~\ref{fig:scaling} shows the normalised
scaling factor $\frac{a'}{a^{1/3}}$ of the final topology ($\chi=1$)
as a function of $\nicefrac{k_{c}}{k_{p}}.$ Low cyclisation to propagation
ratios lead to topologies that possess a scaling factor under critical
level, which implies that they are too dense to become embedded in
a three-dimensional space.

\paragraph{Timeline of network topology}

\begin{figure}
\center \includegraphics[width=1.1\textwidth]{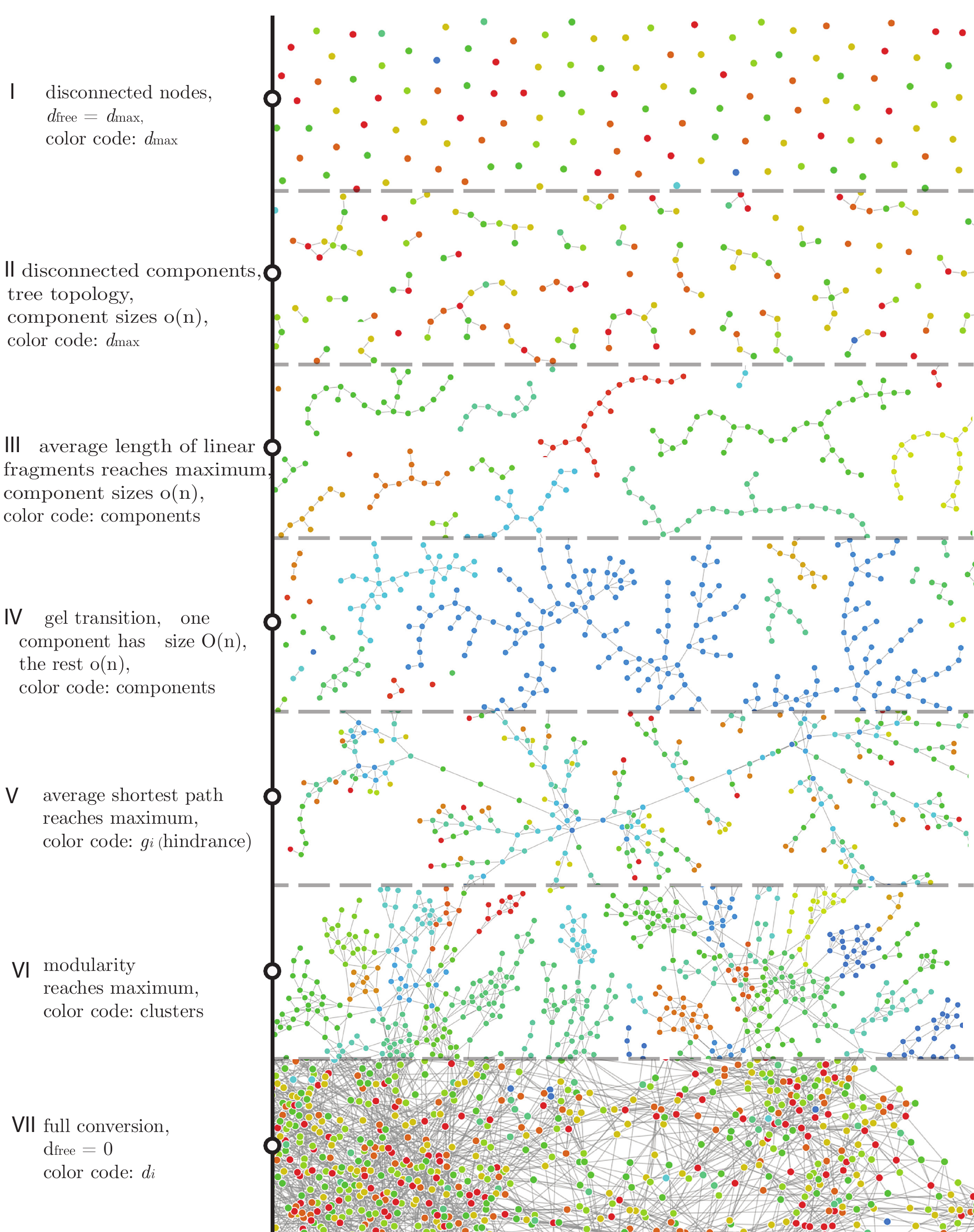} \caption{Timeline of network assembly, $k_{c}/k_{p}=10$ (weak cyclisation)}
\label{fig:timelineA} 
\end{figure}

\begin{figure}
\center \includegraphics[width=1.1\textwidth]{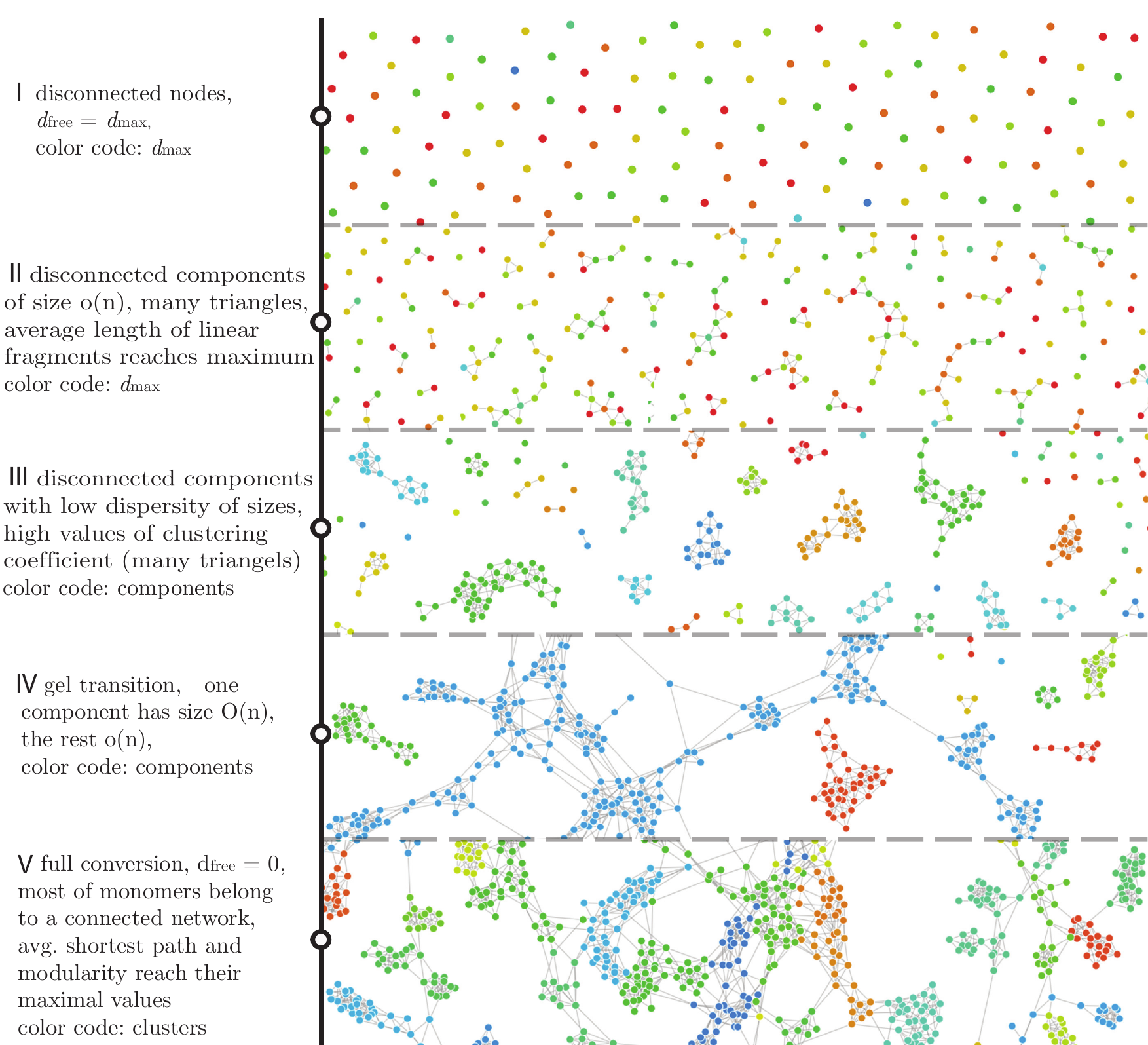} \caption{Timeline of network assembly, $k_{c}/k_{p}=300$ (strong cyclisation)}
\label{fig:timelineB} 
\end{figure}

Even though the final degree distribution of the monomers is fixed,
various input parameters yield a vast range of different network topologies.
We will now comment on the complex process assembling these topologies
by highlighting a few important stages. The input parameters influence
the positions of these stages in time/conversion but the order remains
unaltered. In all cases the polymerisation obviously starts with all
nodes disconnected, $d_{free}=d_{max}$ (see Figures \ref{fig:timelineA}-I
and \ref{fig:timelineB}-II). We now separately consider scenarios
for weak ($k_{c}/k_{p}$ is small) and strong ($k_{c}/k_{p}$ is high)
cyclisatios. When cyclisation is weak, shortly after the start of
the polymerisation, many small disconnected components with tree-topology
are formed (Figure~\ref{fig:timelineA}-II). The sizes of the components
increase but remain $o(n)$; the average length of the linear fragments
continues to increase until it reaches the global maximum (Figure~\ref{fig:timelineA}-III).
The giant component of size $O(n)$ emerges, while cycles start to
appear in all components (Figure~\ref{fig:timelineA}-IV). These
cycles remain being short and consequently do not considerably affect
the average shortest path length. As the network is becoming more
dense, long cycles start to appear occasionally. These long cycles seem
to force the average shortest path length to start declining (Figure~\ref{fig:timelineA}-V).
As the density of the network continues to increase, modularity reaches
its maximal value: clear structural non-homogeneity is present. (Figure~\ref{fig:timelineA}-VI).
At the full conversion the network consist of a single connected component.
The last reactions, even at very low rates, connect the previously
formed clusters and thus reduce the non-homogeneity. The final network
is quite dense and homogeneous (Figure~\ref{fig:timelineA}-VII).

Whereas in the case of strong cyclisation, small molecules with
many cycles form shortly after the start of the reaction (Figure~\ref{fig:timelineB}-II).
The disconnected components grow in size for a much longer span of
conversion before gel transition is achieved. Most of the new edges
appear inside the connected components, enhancing the clustering coefficient
(Figure~\ref{fig:timelineB}-III). The giant component emerges at
a relatively high conversion and it has a non-homogenous structure.
(Figure~\ref{fig:timelineB}-IV). Finally, at the full conversion the
major part of the network is occupied by the giant component, although
other components of small size, $o(n)$, remain present too. The structure
is highly non-homogenous. The full conversion is also the point at which
the modularity and the average shortest path length reach maximum
values. 

\paragraph{Role of excluded volume principle}

An important and distinct feature of the RGM approach is that reaction
rates depend on distances between monomers that are not known directly,
but are implicitly inferred from the topological configuration of
the network. In order to define the connection between topology and
physical distance we employ the 'random walk' concept. The question
arises, which of the random-walk models should be chosen: the commonly used random coil model, $\Phi^{*}$ as given
in \eqref{eq:32}, or the self-avoiding random walk (excluded volume
principle $\Phi,$ Equation \ref{eq:Phi})? A further question is:
how would accuracy be improved in the final results by adopting the
more realistic model? It turns out that the choice between the two
models is not just a matter of accuracy, since using
the random coil model would lead to completely erroneous results for
large networks. The reason behind this phenomenon is the asymptotic
behaviour of the models on increasingly large distance. 

Let's consider a connected infinite network of monomers being homogeneously
distributed in space. Two spheres of radii $p,p+\delta,$ are centred
at a selected monomer. The number of monomers positioned in the volume
between the two spheres is proportional to $\delta p^{2}$ when $\delta\rightarrow0$.
Thus, for any small $\delta>0,$ the asymptotical behaviour of the
reaction probability is given by, 
\[
\begin{aligned}\label{eq:ref}\lim\limits _{p\rightarrow\infty}\delta p^{2}\Phi^{*}(A) & =\infty,\\
\lim\limits _{p\rightarrow\infty}\delta p^{2}\Phi(A) & =0.
\end{aligned}
\]
In other words, in a model equipped with $\Phi^{*}(A)$ (unlike $\Phi(A)$)
the reaction probability increases for increasing distance between
the reacting monomers. This eliminates the random coil model
$\Phi^{*}$ as a proper candidate to be employed in simulations of
infinite network.

\section{Conclusions}

We designed a new stochastic approach for modelling molecular networks
that lies at the intersection of two disciplines: polymer reaction
engineering and network science. The kinetics is accounted for by
a Markov Chain accelerated Gillespie Monte Carlo routine. The main
output is the network topology as presented by an adjacency matrix.
Utilising adjacency matrices allows analyses with the whole ensemble
of tools coming from graph theory: average shortest path, clustering,
modularity, giant component transition, etc. These results are then
used to estimate how the spatial position of monomers incorporated
in the network affects their reactivity. The model turns out to reproduce
polymer (network) characteristics like gel point transition, molecular
size distributions, length of linear fragments. Additionally, the
model gives insight into nuances of topological structure, for instance:
the measure of non-homogeneity or cluster (micro-gel) distribution.

We have successfully applied the RGM algorithm to linseed oil polymerisation,
more commonly known as the drying of oil paint. This process produces
a TAG network that involves monomers of various functionality up 9,
forming the building blocks the network. We showed that even though
the degree distribution is fixed, as it is dictated by the natural
abundance of fatty acids of different unsaturation in linseed oil,
a wide range of different topologies emerge. Strong cyclisation (caused
for instance by diluting the system) leads to highly non-homogeneous
networks that are also less robust. In contrast, weak cyclisation
causes more robust and homogenous structures.

If intra-monomeric reactions, that reduce functionality
of monomers, take place even at a very slow rate they will considerably contribute
to the final structure of the network in the post-gel regime, when rates of all other competing reactions are slowed down.

The resulting network topology is presented as an adjacency matrix,
which allows employing graph theory oriented mechanical models
to study elastic, transport, or rheological properties of the materials.

The main outcome of this work is an algorithm generating molecular
topologies. However, alongside we have introduced a new vocabulary
inspiring a new way of discussing and comparing random molecular topologies,
even regardless of the the methodology adopted to construct them.

\section*{Acknowledgement}

This study has been funded by the PAinT (Paint Alterations in Time)
project as part of the NWO (Dutch Science Foundation) Science4Arts
Program. The authors (Jorien, Ivan, Joen, Piet) thank dr.
Katrien Keune (Rijksmuseum/UvA), dr. Annelies van Loon (UvA), dr.
Maartje Witlox-Stols (UvA), and Marta Mouro, MSc (UvA) for the inspiring
discussions that have led to the problem formulation.
\bibliographystyle{plain}
\bibliography{literature}

\end{document}